\documentclass[12pt,preprint]{aastex}
\usepackage{amsmath}
\usepackage{amssymb}
\usepackage{graphicx}
\usepackage{subfigure}
\usepackage{color} % For use only with the \remark{} macro below.

\slugcomment{Nov. 13th, 2010}
\newcommand{\bfnabla}{{\mbox{\boldmath $\nabla$}}}

\newcommand{\QT}{Q_{\rm T}}
\newcommand{\RH}{R_{\rm H}}
\newcommand{\rS}{r_{\rm S}}

\newcommand{\msun}{M_\odot}
\newcommand{\mbh}{M_{\rm{BH}}}
\newcommand{\del}{\partial}
\renewcommand{\bv}{{\mbox{\boldmath $v$}}}
\newcommand{\avg}[1]{\langle\langle#1\rangle\rangle}
\newcommand{\solaryr}{M_{\odot}\ \mbox{yr}^{-1}}
\shortauthors{Y.-F. Jiang \& Jeremy Goodman} \shorttitle{Quasar
Disk}

\begin{document}

\title{Star Formation in Quasar Disk}

\author{Yan-Fei Jiang\altaffilmark{1}  \& Jeremy Goodman\altaffilmark{1}  }
\affil{$^1$Department of Astrophysical Sciences, Princeton
University, Princeton, NJ 08544, USA}
\begin{abstract}
  Using a version of the ZEUS code, we carry out two-dimensional
  simulations of self-gravitating shearing sheets, with application to
  QSO accretion disks at a few thousand Schwarzschild radii,
  corresponding to a few hundredths of a parsec for a
  100-million-solar-mass black hole.  Radiation pressure and optically
  thick radiative cooling are implemented via vertical averages.  We
  determine dimensionless versions of the maximum surface density,
  accretion rate, and effective viscosity that can be sustained by
  density-wave turbulence without fragmentation. Where fragments do
  form, we study the final masses that result.  The maximum
  Shakura-Sunyaev viscosity parameter is approximately $0.4$.
  Fragmentation occurs when the cooling time is less than about twice
  the shearing time, as found by Gammie and others, but can also occur
  at very long cooling times in sheets that are strongly
  radiation-pressure dominated.  For accretion at the Eddington rate
  onto a $10^8$ solar-mass black hole, fragmentation occurs beyond
  four thousand Schwarzschild radii, $\rS$.  Near this radius, initial
  fragment masses are several hundred suns, consistent with estimates
  from linear stability; final masses after merging increase with the
  size of the sheet, reaching several thousand suns in our largest
  simulations.  With increasing black-hole mass at a fixed Eddington
  ratio, self-gravity prevails to smaller multiples of $\rS$, where
  radiation pressure is more important and the cooling time is longer
  compared to the dynamical time; nevertheless, fragmentation can
  occur and produces larger initial fragment masses.  Because the
  internal thermal and gravitational energies of these massive,
  radiation-pressure-dominated fragments nearly cancel, small errors
  in energy conservation can cause spurious results such as
  spontaneous dissolution of isolated bodies, unless special care is
  taken.  This is likely to be a challenge for all eulerian codes in
  self-gravitating regimes where radiation pressure dominates.

\end{abstract}

\keywords{ accretion, accretion disks --- galaxies: active---
gravitation---  methods: numerical --- stars: formation }

\section{Introduction}
\label{sec:intro}
Active galactic nuclei (AGN) are powered by accretion onto a
supermassive black hole (SMBH).  Probably but less certainly,
accretion occurs via gaseous disks, and furthermore via a local
balance between inward advection of angular momentum and outward
transport by torques due to magnetohydrodynamic turbulence, spiral
density waves, or magnetized winds.  Though mostly indirect, the
evidence for disks at distances $\lesssim 1$~pc from the SMBH is
extensive.  Theoretically, disks naturally result from the combination
of energy dissipation and angular-momentum conservation; also, disk
accretion down to the marginally stable orbit naturally accounts for
conversion of $\sim 10\%$ of accreted mass to radiation, as required
by comparisons of the integrated light from AGN with the integrated
mass in SMBHs (\citealt{Soltan1982}; \citealt{YuTremaine2002}).
Phenomenologically, disk accretion is consistent with the evidence for
anisotropic emission as embodied in the unification model for AGN
\citep{MillerAntonucci1983}.  In a few highly selected AGN, 
VLBI observations of maser emission strongly indicate the presence of a
gaseous disk in keplerian rotation at a few tenths of parsecs from
the SMBH \citep{Miyoshi_etal1995,Kuo_etal2010}.

A longstanding difficulty with the disk paradigm for AGN is to explain
how gas is transported from galactic to sub-parsec scales without
turning entirely into stars \citep{Shlosmanetal1990}.  Disk accretion
in local thermal equilibrium is prone to self-gravity because the
vertical thickness tends to be small and the inflow speed slow, so
that the density of the gas must be large to support inferred mass
accretion rates.  On scales $\gtrsim GM_{BH}/\sigma_{\rm
  circ,gal}^2\sim 10$~pc, external gravitational torques due to
merging galactic nuclei or stellar bars may speed the inflow.  On
scales $\lesssim 10^3 GM_{BH}/c^2\sim 10^{-2}$~pc, the tidal field of
the SMBH is large enough and the temperature of a standard viscous
disk is high enough so that self-gravity is typically slight.  This
leaves a broad range of radii, typically $0.01$-$10$~pc or
$10^3$-$10^6$ Schwarzschild radii $r_{\rm S}$, over which a standard
viscous disk in steady-state accretion would be self-gravitating, with
a Toomre $Q$ parameter that decreases rapidly with increasing radius
\citep{Goodman2003}.  Gravitational fragmentation, leading to star
formation within the disk, appears to be a natural outcome.  Indeed,
stellar disks are common within a few parsecs of the SMBH in nearby
quiescent galaxies \citep{Laueretal2005} and active Seyferts
\citep{Daviesetal2007}, while our own Galactic Center contains what
appear to be main-sequence B~stars at $\sim 10^{-2}$~pc
\citep{Ghez_etal2003,Martins_etal2008}.

There is a general belief that star formation and accretion somehow
regulate and support one another in AGN (e.g.,
\citealt{CollinZahn1993,ThompsonQuataertMurray2005}).  But, we
believe, a convincing model of how this works is still lacking.
Beyond the usual perplexities attending ``normal'' star-formation,
sub-parsec AGN disks pose severe problems of energetics and stability.
Specific orbital energies are not small compared to those released by
stellar evolution ($\sim 10^{-3}c^2$), while orbital and thermal
timescales are much shorter than the minimum main-sequence lifetime
($\sim10^6$~yr); both comparisons would appear to make a stable
feedback between gravitational collapse and stellar energy inputs more
difficult than in giant molecular clouds.  Also, it is not clear that
AGN disks should form stars of normal mass.  Evidence exists for a
somewhat top-heavy stellar mass function among the young stars in the
Galactic Center \citep{NayakshinSunyaev2005,Bartko_etal2010}.
\cite{GoodmanTan2004} argued that objects of up to $\sim
10^5\,\msun$ might form, this being the so-called ``isolation mass''
(a dynamical scale borrowed from theories of planet formation)
appropriate to radii $\sim 10^3\,r_{\rm S}$ in a typical bright AGN.

In view of the physical and observational difficulties of this
subject, a complete theory may not be available for many years, but
progress has been made in understanding the role of self-gravity in
promoting accretion, and the conditions for fragmentation.
\cite{Gammie2001} showed via idealized two-dimensional simulations
that disks (or rather, shearing sheets) subject to cooling can
maintain themselves in a state of marginal linear stability according
to the \cite{Toomre1964} criterion provided that the product $\Omega
t_c$ cooling time and orbital angular velocity is somewhat greater
than unity, where $t_c$ is the timescale on which the gas would
radiate its thermal energy in the absence of heating processes.
Gammie's models with $\Omega t_c\gtrsim 1$ reached a statistical
steady state with nonlinear density waves and shocks sufficient to
offset the cooling.  Since the energy source that supports ongoing
mechanical dissipation is ultimately differential rotation, the wave
turbulence must transport angular momentum outward, and since Gammie's
models were effectively local this transport can be described by a
Shakura-Sunyaev viscosity parameter $\alpha\approx (\Omega t_c)^{-1}$,
the exact value depending upon the assumed ratio $\gamma$ of specific
heats of the gas.  Below a critical value $\Omega t_c\approx 3$,
Gammie's disk fragmented into gravitationally bound objects.  His
conclusions have been confirmed, with some variations in the critical
values of $\Omega t_c$ and $\alpha$, by subsequent simulations with
more realistic cooling \citep{JohnsonGammie2003} and with global,
three-dimensional geometries \citep{RiceArmitageBonnell2003,LodatoRice2004}.

Recently, \cite[hereafter HQ]{HopkinsQuataert2010a,HopkinsQuataert2010b}
have argued, based on SPH simulations buttressed by analytic arguments, that global,
nonlinear, non-axisymmetric density waves accompanied by shocks can
support accretion rates as high as $\sim10\solaryr$ at $\lesssim
0.1$~pc.  One-armed spirals (azimuthal mode number $m=1$) are
particularly prominent in their simulations, probably because such
modes cohere most easily in near-keplerian potentials (e.g.,
\citealt{LeeGoodman1999} and references therein).  It is true that
nonlinear global spirals can in principle exert much larger torques on
the gas than is possible with a local effective viscosity, by a factor
$\sim (r/h)\alpha^{-1}$, where $h$ is the disk thickness and $\alpha$
is the Shakura-Sunyaev viscosity parameter \citep[hereafter
G03]{Goodman2003}.  HQ's results sidestep rather than solve the
problem of local self-gravity, however.  Their simulations employ a
superthermal effective sound speed that is intended to represent
unresolved turbulence, and which is large enough to suppress local
instability.  This device was originally developed to parametrize
stellar feedback on galactic scales
\citep{SpringelDiMatteoHernquist2005}, but for the reasons of
energetics and timescales mentioned above, one may question its
applicability to the scales of interest to us, $0.01$-$0.1$~pc from
the SMBH.  In any case, global waves, particularly HQ's
near-stationary $m=1$ waves, could not be observed in shearing-sheet
simulations such as those of our paper.

In all the simulations following \cite{Gammie2001}'s original work,
one issue that has not been directly addressed is the role of
radiation pressure in the equation of state, perhaps because most of
the applications have been to protostellar disks or to the Galactic
Center.  For accretion rates and black-hole masses characteristic of
bright AGN, however, radiation pressure is still important at the
minimum radius where self-gravity sets in (G03).  As is well known,
the ratio of radiation pressure to gas pressure is a monotonically
increasing function of mass for optically thick, nondegenerate bound
objects in hydrostatic equilibrium; and the effective adiabatic index
$\Gamma_1=(\partial\ln p/\partial\ln\rho)_S$ a correspondingly
decreasing function.  This suggests that as objects gain mass through
accretion or merging in a fragmenting AGN disk, they become
increasingly susceptible to rapid collapse.  As noted by
\cite{Gammie2001} and confirmed in three dimensions by \cite{RiceLodatoArmitage2005},
the critical value of $\Omega t_c$ for
fragmentation depends upon the effective equation of
state relating pressure ($p$) to mass density ($\rho$), or
height-integrated pressure, $P$, to surface mass density, $\Sigma$.
In particular, if $\gamma_{2D}=(\partial\ln
P/\partial\ln\Sigma)_S\le3/2$, then collapse may occur without any
cooling ($\Omega t_c=\infty$).  This perhaps is why Gammie chose to do
his simulations with $\gamma_{2D}=2$.  The correspondence between
$\Gamma_1$ and $\gamma_{2D}$ depends upon how the vertical thickness
of the disk responds to changes in surface density: if the response is
hydrostatic and strongly self-gravitating, then $\gamma_{2D}=3/2$
corresponds to $\Gamma_1=4/3$, the value for a spherical body
supported entirely by radiation pressure.  

Most of these points regarding the importance of radiation pressure were
made by GO3 and \cite{GoodmanTan2004} via analytical arguments; the principal
goal of the present work is to illustrate them through numerical
simulations.  We originally hoped also to demonstrate growth up to the
isolation mass (appropriately redefined for the
shearing sheet), but while we do demonstrate growth well beyond the
mass scale associated with linear instability (``Toomre mass''),
numerical difficulties described below prevented us from making a
systematic study of the ultimate masses as a function of
shearing-sheet control parameters.

The structure of this paper is as follows. In \S\ref{sec:EOS}, we
present the adopted equation of state for our height-integrated,
two-dimensional calculations; some mathematical details are deferred
to an Appendix.  Our equation of state incorporates both radiation
pressure and self-gravity under the assumption of local vertical
hydrostatic equilibrium.  In \S\ref{sec:equations}, we review the
basic equations for the shearing sheet.  Our cooling prescription,
which is based on a simple algebraic approximation to vertical
radiative transfer, is presented in \S\ref{sec:cooling}.  We describe
our computational units and control parameters in \S\ref{sec:units},
and our numerical methods in \S\ref{sec:numerical}. In
\S\ref{sec:Results}, we present results from representative
simulations on both sides of the fragmentation boundary.  A general
picture of the disk based on our simulations is described in
\S\ref{Results:Generalpicture}.    A summary and discussion of our
main results, and some speculations concerning observable
consequences, are given in \S\ref{sec:discussion}.

\section{A shearing-sheet model for AGN disks}
This section presents the height-integrated physical model upon which
our numerical simulations are based.

\subsection{The equation of state}
\label{sec:EOS}

Let $(r,\phi, z)$ be cylindrical coordinates such that the central
black hole, with Schwarzschild radius $r_s=2G\mbh/c^2$,
lies at $r=0$, and the disk midplane at $z=0$.
At radii $r\sim 10^3$-$10^4\,\rS$, the disk is expected to be quite
thin, with a half thickness $h\sim 10^{-2}r$ (G03).
The $z$ coordinate is therefore eliminated from the governing  equations of our
numerical simulations by vertical integration so that, for example, pressure
$p$ and mass density $\rho$ are replaced by height-integrated pressure $P$
and surface density $\Sigma$:
\begin{equation}
P(r,\phi,t)=\int_{-\infty}^{+\infty}p(z,r,\phi,t)dz,\qquad\qquad
\Sigma(r,\phi,t)=\int_{-\infty}^{+\infty}\rho(z,r,\phi,t)dz.
\end{equation}

By an ``equation of state,'' we mean a functional relation among
$P$, $\Sigma$, and other thermodynamic variables.
To obtain such a relation, we make a
number of simplifying assumptions about the thermodynamics of the gas
and about the distribution of $p$ and $\rho$ with $z$.
Since the regions of the disk with which we are concerned are probably
very optically thick, the gas and radiation temperature are taken equal.  
Vertical hydrostatic equilibrium is assumed, and
magnetic and turbulent contributions to the $zz$ component of
the stress tensor are neglected, so that the disk thickness is supported 
entirely by the sum of gas and radiation pressure, $p=p_r+p_g$.  The
gas pressure fraction
\begin{equation}\label{beta}
\beta \equiv \frac{p_g}{p_g+p_r}
\end{equation}
is presumed constant with height but allowed to vary with
$(r,\phi,t)$.   In combination with vertical hydrostatic equilibrium,
constant $\beta$ implies constant $\kappa F_z/g_z$, where $\kappa$
is the opacity; $g_z$ is the vertical component of the gravitational
field; and $F_z$ is the vertical component of the radiative heat flux.
The vertical runs of density and pressure 
are then related by a polytropic relation,
\begin{equation} \label{pressureK}
p=K(\beta)\rho^{4/3}, \qquad K(\beta)=\left[\frac{3}{a}\left(\frac{k_B}{\mu
m_p}\right)^4\frac{1-\beta}{\beta^4}\right]^{1/3}\,,
\end{equation}
with molecular weight $\mu\approx 0.62$,
as appropriate for a fully ionized gas of near-solar metallicity.

Once $\beta$ is specified, $P$ can be found in terms of $\Sigma$ by
inserting eq.~\eqref{pressureK} into the equation of vertical
hydrostatic equilibrium.  But when self-gravity is included [via the
one-dimensional form \eqref{verticalbalance} of Poisson's equation],
the relationship that results is implicit.  We relegate the
mathematical details to the Appendix and simply quote the main results
here.

It is convenient to introduce the quantity
\begin{equation} \label{Q}
 Q\equiv\frac{\Omega^2}{2\pi G\rho(0)},
\end{equation}
where $\rho(0)$, which is shorthand for $\rho(0,r,\phi,t)$, is the
mass density at the midplane, and $\Omega=(G\mbh/r^3)^{1/2}$ is the
orbital angular velocity.  We use this notation because $Q$
as we have defined it is usually numerically comparable to Toomre's
stability parameter $\QT\equiv\Omega C_s/\pi G\Sigma$, to which it
would reduce if the effective thickness $\Sigma/\rho(0)$ of the disk
were given by $2 C_s/\Omega$ in terms of the effective
\emph{horizontal} sound speed $C_s\equiv(\partial
P/\partial\Sigma)_S^{1/2}$.  The actual value of the effective
thickness is somewhat different from $2 C_s/\Omega$, however, not only
because of vertical variations in the true three-dimensional sound
speed $c_s\equiv(\partial p/\partial\rho)_S^{1/2}$, but also because
of the self-gravity of the disk.  The quantity \eqref{Q} is more
directly related to the Roche criterion for an object of mean density $\sim\rho(0)$.

It is shown in the Appendix that
\begin{equation}
P=\frac{\pi^2QI_4(Q)}{16[I_3(Q)]^{3}}\frac{G^2\Sigma^3}{\Omega^2}\,,
\label{EOS}
\end{equation}
\begin{equation}
K(\beta)=\frac{(\pi
G)^{7/3}}{2^{11/3}}\frac{Q^{4/3}\Sigma^2}{\Omega^{8/3}\left[I_3(Q)\right]^{2}}\,,
\label{betaandQ}
\end{equation}
and
\begin{equation}
\beta=2-\frac{32\left[I_3(Q)\right]^3}{3\pi^2I_4(Q)Q}\frac{\Omega^2U}{G^2\Sigma^3}\,,
\label{EOS2}
\end{equation}
in which
\begin{equation}
I_3(Q)\approx\frac{0.323}{\sqrt{Q+1.72}},\qquad\qquad
I_4(Q)\approx\frac{0.287}{\sqrt{Q+1.72}} \,.\label{I3I4}
\end{equation}
Equation \eqref{EOS} determines $P$ in terms of
$\Sigma$ and $Q$, but $Q$ is not a conserved or primitive variable in
our dynamical equations.  Instead, we have the thermodynamic internal
energy per unit area, $U$.  Since the internal energy per unit volume
$u=(3/2)p_g+3p_r= 3(1-\beta/2)p$ under our assumptions of complete
ionization and equal gas and radiation temperatures, and since $\beta$
is assumed independent of $z$, it follows that
\begin{equation}
  \label{UvsP}
  U=\left(1-\frac{\beta}{2}\right)P.
\end{equation}
Since $K(\beta)$ is the function \eqref{pressureK}, equations
\eqref{betaandQ} and \eqref{EOS2} determine $\beta$ and $Q$ in terms
of $\Sigma$, $U$, and $\Omega$, and then eq.~\eqref{UvsP} gives $P$.
It can be shown from eqs.~\eqref{EOS}, \eqref{EOS2}, and \eqref{I3I4}
that $P\propto\Sigma^{3/2}$ when both $\beta$ and $Q$ are $\ll 1$,
i.e. the effective 2D adiabatic index approaches the critical value of
$3/2$ at which nonrotating bound fragments can collapse indefinitely without cooling.

For the calculation of the local cooling time, we sometimes require
the physical density and temperature since we include a Kramers
component in our opacity law (\S\ref{sec:equations}).
The mid-plane density $\rho(0)$ follows from $Q$ via eq.~(\ref{Q}), and
the mid-plane temperature $T(0)$ is
\begin{equation}
T(0)=\left[\frac{3k_B}{a\mu
m_p}\frac{1-\beta}{\beta}\rho(0)\right]^{1/3}. \label{midT}
\end{equation}

\subsection{Dynamical equations}\label{sec:equations}
Following \cite{Gammie2001}, we describe the local dynamics of the disk in
a shearing sheet approximation, in which $x=r-r_0$ and
$y=r_0[\phi-\Omega(r_0) t]$ are pseudo-Cartesian coordinates centered on 
a circular orbit of radius $r_0$.  The equations of motion are
\begin{equation}
\frac{\del \Sigma}{\del t}+\bfnabla\cdot(\Sigma\bv)=0,
\label{equationDensity}
\end{equation}
\begin{equation}
\frac{D\bv}{Dt}=-\frac{\bfnabla
P}{\Sigma}-2\Omega\mathbf{e}_z\times\bv+3\Omega^2x\mathbf{e}_x-\bfnabla\Phi,
\label{equationV}
\end{equation}
\begin{equation}
{\frac{\del U}{\del t} + \bfnabla \cdot (U \bv) = -P\bfnabla\cdot\bv
-\Lambda}. \label{equationU}
\end{equation}
\begin{equation}
\nabla^2\Phi=4\pi G\Sigma\delta(z). \label{equationpotential}
\end{equation}
The cooling function $\Lambda$ represents
radiative losses from the surface of the disk, as described
in \S\ref{sec:cooling}.
If $\Lambda=0$, these equations have steady solutions in which
$\Sigma$ and $\Phi$ are constants and
$\mathbf{v}=-\tfrac{3}{2}\Omega x\mathbf{e}_y$.

A useful diagnostic quantity is the potential vorticity
\begin{equation}
\xi\equiv \frac{\bfnabla \times \bv+2\mathbf{\Omega}}{\Sigma}.
\label{xi}
\end{equation}
This is conserved following the fluid, $D\xi/Dt=0$, in the absence of
shocks, viscosity, or uneven cooling, so that the equation of
state is effectively barotropic, $P=P(\Sigma)$.

% If we include magnetic field, then there will be additional
% viscosity due to magnetorotational  instability (MRI) (e,g.,
% \citealt{BalbusHawley1991}). If the MRI effect can be parametrized
% by $\alpha_m$ and dynamic viscosity coefficient due to MRI is
% $\eta_m=\alpha_m P/\Omega$. Then with this additional viscosity
% term, the equations for the velocity (\ref{equationV}) and internal
% energy (\ref{equationU}) are changed to be
% \begin{equation}
% \frac{D\bv}{Dt}=-\frac{\bfnabla
% P}{\Sigma}-2\bold{\Omega}(r_0)\times\bv+3\Omega^2x\bold{e}_x-\bfnabla\Phi
% -\frac{\bfnabla\cdot\mathbf{\Pi}}{\Sigma}, \label{equationV2}
% \end{equation}
% \begin{equation}
% \frac{\del U}{\del t} + \bfnabla \cdot (U \bv) = -P\bfnabla\cdot\bv
% -\Lambda+\Psi. \label{equationU2}
% \end{equation}
% Here $\mathbf{\Pi}$ is the viscosity tensor given by
% \begin{equation}
% \mathbf{\Pi}=-\eta_m\left(
% \begin{array}{cc}
%  \frac{\del v_x}{\del x}-\frac{\del v_y}{\del y} & \frac{\del v_y}{\del x}+\frac{\del v_x}{\del y}  \\
%  \frac{\del v_x}{\del y}+\frac{\del v_y}{\del x} & \frac{\del v_y}{\del y}-\frac{\del v_x}{\del x}
% \end{array}
% \right),
% \end{equation}
% while $\Psi$ is the heating function given by
% \begin{equation}
% \Psi=\sum_{i=1}^2\sum_{j=1}^2\frac{\Pi_{ij}^2}{2\eta_m},
% \end{equation}
% where $\Pi_{ij}$ is the element of the tensor $\mathbf{\Pi}$ at $i$
% line and $j$ column.
% 

We do not include any explicit viscous terms.  Since we do not resolve
the vertical dimension of the disk/sheet, we could not represent
magnetororational instabilities (MRI) directly but would have to
parametrize the magnetic stresses in terms of $\Sigma$ and $U$; such
parametrizations are prone to thermal and viscous instabilities where
radiation pressure dominates
\citep{LightmanEardley1974,HiroseBlaesKrolik2009}.  Furthermore, the
effective Shakura-Sunyaev parameter provided by density waves and
shocks in our simulations is usually larger than
$\alpha_{MRI}\sim 10^{-2}$.  There is, however, an
artificial viscosity included in ZEUS to mediate shocks (e.g.,
\citealt{StoneNorman1992}).

\subsection{Cooling Fuction}\label{sec:cooling}
Following \cite{JohnsonGammie2003}, 
the cooling function $\Lambda$ in equation (\ref{equationU})
represents radiation losses from the surface of the disk.
Since we do not resolve the disk thickness
we adopt a standard algebraic approximation for the vertical radiative transfer:
\begin{equation} \label{cooling0}
\Lambda=2\sigma T_{\rm eff}^4\approx\frac{16}{3}\sigma T^4(0)\,\left(\tau+\frac{1}{\tau}\right)^{-1}\,,
\end{equation}
in which
\begin{equation}
\tau\equiv \frac{1}{2}\Sigma\kappa[\rho(0),T(0)]. \label{tau}
\end{equation}
approximates the local optical depth from the midplane to the surface.
Usually $\tau\gg 1$ on average in those parts of AGN disks that we
wish to model, but in case fragmentation should lead to gaps in the
sheet, the form of equation \eqref{cooling0} is chosen so that
$\Lambda\propto\tau$ in the optically thin limit
(e.g. \citealt{Hubeny1990,JohnsonGammie2003}).

At radii $10^3$-$10^4\,\rS$,
the mid-plane temperature $T(0)$ is
typically $10^4$-$10^5\ K$ for near-Eddington accretion rates (GO03). If we assume
$Q=1$, the mid-plane density will be $\rho(0)\sim 10^{-9}{\,\rm g\,cm^{-3}}$.
In this density and temperature range, the dominant
opacity is electron scattering, which is almost constant, justifying
the factor $1/2$ in equation \eqref{tau}.
We include a Kramers opacity, however, which often begins to be important beyond
$\sim 5000\rS$ in our simulations.  An analytic approximation
to the opacity sufficient for our purposes is therefore
% \begin{equation}
% \kappa_e(z)=0.2(1+X)\left[1+2.7\times
% 10^{11}\frac{\rho_z}{T_z^2}\right]^{-1}\left[1+\left(\frac{T_z}{4.5\times10^8}\right)^{0.86}\right]^{-1},
% \label{kappeequa}
% \end{equation}
% and Kramers formula $\kappa_k(z)$ is
\begin{equation}\label{opacity}
\kappa=\kappa_{\rm es}+ \kappa_{\rm K}=0.2(1+X)+4\times10^{25}\,(1+X)(Z+0.001)\frac{\rho}{T^{3.5}}\,,
 \end{equation}
all quantities being evaluated in in cgs units (B. Paczynski, private communication).  The
 mass fractions of hydrogen and metals are taken at their solar
 values, $X=0.7, Z=0.02$.  There is evidence that the broad-line gas
 in quasar stellar objects is more metal-rich than the sun
 (e.g. \citealt{HamannFerland1993,Dhanda_etal2007}); however, this
 would not much affect our results because electron scattering opacity
 dominates as long as the metallicity of the broad-line gas is smaller
 than $\sim 10$ of the solar value, at least for near-Eddington
 accretion onto black holes of masses $\gtrsim 10^8 \msun$ in the
 range of radii where self-gravity is important but fragmentation is avoided.
Kramer's opacity gains in importance with increasing radius,
decreasing $\mbh$,  and decreasing $\dot M$.

The local cooling time is defined as
\begin{equation}
t_c\equiv \frac{U}{\Lambda}.
\end{equation}
For $Q\gtrsim 1$, the half-thickness $h\sim\sqrt{P/\Sigma}$.  It
follows that $t_c\sim \kappa T^4(0)/(c\Omega)^2$ for $\beta\ll 1$, 
and that $t_c\propto\kappa \Sigma^2/T^3(0)$ when $\beta\approx 1$.

\subsection{Computational units} \label{sec:units}
In our simulations, it is convenient to scale the
fluid variables so that they are of order unity.
To this end, we adopt the time unit
\begin{equation}
t_0\equiv\Omega^{-1}\approx1.4M_8r_3^{3/2}\ \mbox{yr}, \label{unitt}
\end{equation}
 where $M_8\equiv \mbh /(10^8M_{\odot})$ and $r_3\equiv r/(10^3r_s)$.
As already noted, this is very short compared to the nuclear timescale
of a massive star, $t_{\rm nuc}=0.007 M_* c^2/L_{\rm Edd}(M_*)\sim
10^6$~yr, and short even compared to the main-sequence
Kelvin-Helmholtz timescale, which approaches $\sim 3000$~yr from above
for very massive stars (e.g., \citealt{BondArnettCarr1984,GoodmanTan2004}).
% \begin{equation}
% t_{\text{KH,ms}}\approx 3300\left(\frac{\kappa}{\text{0.4 g cm$^{-2}$}}\right)
% \left(\frac{\mu_{\odot}}{\mu}\right)^2\beta^{-0.053}(1-\beta)^{-0.89} \text{ yr}.
% \end{equation}
%If $\beta=0.5$, then $ t_{\text{KH,ms}}\approx 6344$ yr, which is
%much larger than the dynamical time.  
It is in part this disparity of timescales that causes us to doubt the
ability of stellar feedback to stabilize the disk.  More relevant to
our simulations is the Kelvin-Helmholtz timescale for a
radiation-pressure-dominated cloud of mass $M_*$ if we scale its radius $R_*$
by its Hill radius \eqref{Hillradius}:
\begin{equation}
 t_{\text{KH}}\approx 50\left(\frac{M_{\star}}{300M_{\odot}}\right)\left(\frac{R_H}{R_{\star}}\right)
 \left(\frac{\kappa}{\text{0.4 g cm$^{-2}$}}\right)\left(\frac{\beta}{1-\beta}\right)\textrm{yr}\,.
 \label{tKH}
 \end{equation}

We choose our mass unit as
\begin{equation}\label{massunit}
M_0\equiv\left(\frac{3}{aG^3}\right)^{1/2}\left(\frac{k_B^2}{\mu ^2
m_p^2}\right)=10.24M_{\odot}\left(\frac{\mu_{\odot}}{\mu}\right)^2,
\end{equation}
Notice that, apart from the molecular weight $\mu$, this is entirely
determined by fundamental constants ($\mu_{\odot}\approx 0.62$ is the
molecular weight of a fully ionized gas at solar abundance).
This choice 
simplifies the Eddington relation between the mass of a
homogeneous self-gravitating sphere and its gas-pressure fraction:
\begin{equation}
M_{\star}\approx
47M_{\odot}\frac{\sqrt{1-\beta}}{\beta^2}\left(\frac{\mu_{\odot}}{\mu}\right)^2.
\label{Mbeta}
\end{equation}
At $M=M_0$, for example, this predicts $\beta=0.96$.  By no
coincidence, $M_0$ is characteristic of a moderately massive star.

Because of the importance of self-gravity, it is convenient to choose
the length unit so that Newton's constant is of order unity.  The choice
\begin{equation}
L_0\equiv \left(\frac{2\pi
GM_0}{\Omega^2}\right)^{1/3}=2.59\times10^{14}\left(\frac{\mu_{\odot}}{\mu}\right)^{2/3}r_3M_8^{2/3}\
\mbox{cm}
\end{equation}
implies $G=(2\pi)^{-1}$.
Then the units for surface density $\Sigma_0$, velocity $v_0$,
internal energy per unit area $U_0$ and 2D pressure $P_0$ are
combinations of $t_0$, $M_0$, and $L_0$:
\begin{equation}\label{Sigma0}
\Sigma_0\equiv
M_0L_0^{-2}=3.05\times10^5\left(\frac{\mu_{\odot}}{\mu}\right)^{2/3}r_3^{-2}M_8^{-4/3}\qquad
\mbox{g\ cm$^{-2}$},
\end{equation}
\begin{equation}
v_0\equiv\frac{L_0}{t_0}=5.75\times10^6\left(\frac{\mu_{\odot}}{\mu}\right)^{2/3}r_3^{-1/2}M_8^{-1/3}\qquad
\mbox{cm\ s$^{-1}$}.
\end{equation}
\begin{equation}
U_0\equiv P_0\equiv
M_0t_0^{-2}=1.0\times10^{19}\left(\frac{\mu}{\mu_\odot}\right)^{-2}r_3^{-3}M_8^{-2}\qquad
\mbox{erg\ cm$^{-2}$}. \label{unitU}
\end{equation}

Note that our length and time units, but not our mass unit, depend on radius.
We might have avoided this by taking $(\kappa_e M_0)^{1/2}$ 
and $\kappa_e^{3/4}M_0^{1/4}(2\pi G)^{-1/2}$ for our length and time
units, respectively, whence our unit of surface density would be $\kappa_e^{-1}$.
At the radii of interest to us in a bright AGN disk, however,
$\Sigma\sim\Sigma_0\gg\kappa_e^{-1}$ [eq.~\eqref{Sigma0}], so $L_0$ is
the more convenient length standard.
Also, $\Omega$ enters more than once into the Euler equation \eqref{equationV}.
A symptom of this is that the Hill radius works out rather simply:
\begin{equation}
R_H\equiv\left(\frac{M_{\star}}{3\mbh}\right)^{1/3}r_0\approx 0.376\left(\frac{M_{\star}}{M_0}\right)^{1/3}L_0.
\label{Hillradius}
\end{equation}
This is approximately the largest size at which a bound fragment of mass
$M_*$ can withstand the tidal field.

With these units, we express our dynamical variables in
dimensionless form:
\begin{equation}\label{dimlessvars}
\hat{\Sigma}\equiv\frac{\Sigma}{\Sigma_0},\qquad
\hat{U}\equiv\frac{U}{U_0},\qquad\hat{P}\equiv\frac{P}{P_0},\qquad
\hat{\bv}\equiv\frac{\bv}{v_0}.
\end{equation}

\subsection{Accretion Rate}
\label{sec:accretion}

As a consequence of the shearing-sheet boundary conditions, the
joint average over time and space of the radial mass flux $\Sigma v_x$ 
can be shown to vanish.  Thus, we cannot expect to measure the mass
accretion rate ($\dot M$) directly in our simulations.
Nevertheless, we can measure $\dot M$ indirectly from energy or
momentum balance.

At large radii in a steady keplerian thin disk, energy balance
is expressed by $\sigma T_{\rm eff}^4=3\dot M\Omega^2/8\pi$ (e.g.,
\citealt{Pringle1981}).  In our simulations, the effective temperature
is a function of local parameters via the cooling function \eqref{cooling0}.
Therefore, one local estimator for $\dot M$ is
\begin{equation}
\dot M_\Lambda=\frac{4\pi\Lambda}{3\Omega^2}\,.
\label{mdot1}
\end{equation}
On the other hand, steady angular momentum balance requires
$\dot M\Omega r^2= \Gamma-\Gamma_0$, where $\Gamma$ is the ``viscous''
torque, and $\Gamma_0$ is a constant that can be neglected at large
radii.  In a thin keplerian disk, this reduces to $\dot M=
3\pi\nu\Sigma$, where $\nu$ is the effective viscosity.
In our self-gravitating shearing sheets, the role of $\Gamma$
is played by $r_0(G_{xy}+H_{xy})$, in which $G_{xy}$ and $H_{xy}$ are
the offdiagonal components of the vertically integrated
gravitational and Reynolds stresses
as defined by \cite{Gammie2001} and \cite{JohnsonGammie2003}.
This agrees with $\dot M_\alpha =3\pi\nu\Sigma$ if we define\footnote{As we use a different
equation of state, our normalization of $\alpha$ differs from that of
\cite{JohnsonGammie2003} by a factor involving the adiabatic index.}
$\nu=\alpha P/\Omega\Sigma$,
\begin{equation} \label{alpha}
 \alpha\equiv\frac{2}{3}\,\frac{G_{xy}+H_{xy}}{P}\,,
\end{equation}
and
\begin{equation}
\dot M_\alpha=\frac{2\pi(G_{xy}+H_{xy})}{\Omega}\,.
\label{mdot2}
\end{equation}

In those simulations that reach a statistical steady state, the
spatiotemporal averages of $\dot M_\Lambda$ and $\dot M_\alpha$ agree,
and consequently
$\langle t_c\rangle_{x,y,t}=\langle\alpha^{-1}\rangle_{x,y,t}\Omega^{-1}$
(e.g., \citealt{Pringle1981}).
We prefer the estimator \eqref{mdot1} in these steady cases because it
fluctuates less than $\dot M_{\alpha}$.
When the sheet fragments into isolated masses that
secularly cool, however, thermal equilibrium does not hold.  Then $\dot
M_\alpha$ is the more reliable estimator,  and the ``dissipation'' of the
mean shear associated with the stresses in eq.~\eqref{alpha} is
balanced by increasing epicyclic motions of the fragments.

\section{Numerical method}
\label{sec:numerical} Our simulations use a modified form of the code
developed by \cite{Gammie2001} and \cite{JohnsonGammie2003}. This is a
self-gravitating hydrodynamic code based on ZEUS
(\citealt{StoneNorman1992}), which is a time-explicit, operator-split,
finite-difference method on a staggered mesh.  Details and tests of
this code are described by \cite{Gammie2001}. We just emphasize some
important points here.

The code uses the standard ``shearing box" boundary conditions (e.g.,
\citealt{HawleyGammieBalbus1995}).
For a rectangular domain of dimensions $L_x\times L_y$,
all fluid variables $f$ satisfy
\begin{equation}\label{shearingperiodic}
f(x,y,t)=f(x,y+L_y,t),\qquad\qquad f(x,y,t)=f(x+L_x,y-\tfrac{3}{2}\Omega t L_x,t)\,,
\end{equation}
except that $v_y(x,y,t)=v_y(x+L_x,y-\tfrac{3}{2}\Omega t
L_x,t)+\tfrac{3}{2}\Omega L_x$.  Poisson's equation is solved by
discrete Fourier transforms.  Mass is
conserved up to round-off error, so the areal average of $\Sigma$ is a
constant with time.  A FARGO-like scheme is used to facilitate the
transport substeps \citep{Masset2000,Gammie2001}.

However, even without cooling, total energy---the sum of kinetic,
internal, gravitational, and tidal energy (the tidal potential
$\phi_T=\tfrac{3}{2}\Omega x^2$) is not conserved.  Part of the reason
is the shearing-periodic boundary condition \eqref{shearingperiodic},
which maintains the mean shear.  The nearest thing to an energy
integral is the Jacobi-like quantity
\begin{equation}
\Gamma=\int d^3x \Sigma
\delta(x)\left(\frac{1}{2}v^2+\frac{U}{\Sigma}+\phi_T+\frac{1}{2}\phi\right)\,
\label{Jacobi}
\end{equation}
but this is not constant unless cooling exactly balances the
dissipation of mechanical energy by the stresses \eqref{alpha} \citep{Gammie2001}.

Furthermore, there are numerical errors.  ZEUS's energy equation is
not written in flux-conservation form; indeed, such a form may not be
possible for a razor-thin sheet whose self-gravity is described by the
three-dimensional Poisson operator \eqref{equationpotential}.  The
velocities and mass densities are updated in such a way that the
changes in kinetic energy due to gravitational forces are slightly
inconsistent with changes in the self-gravity itself.  The errors are
second-order in space but only first-order in time.  When density
fluctuations are well resolved by the grid, these errors are minor:
typically less than 10\% over several hundred dynamical times.  For
compact fragments that span only a few cells, however, the fractional
error in the binding energy after several shearing times can become
more than 100\%.  Simulations of Jeans collapse in nonshearing
($\Omega\to0$) sheets exhibit the problem clearly when the Jeans
length is comparable to the grid scale.  Massive ($M\gg M_0$)
fragments are especially problematic with our ``soft'' equation of
state, because thermal and gravitational energies nearly cancel for a
nonrotating bound object that is radiation-pressure dominated
($\beta\ll 1$).  Because \cite{Gammie2001} and
\cite{JohnsonGammie2003} use a relatively ``hard'' equation of state,
the error is less important, although it may contribute to the small
deviations from the expected relation in their Figure~12.
% \footnote{We want to point out that there is
% another bug in the transport step of the original code, which is
% corrected here. The original code calculates the transport of the
% velocity along $y$ direction slightly wrong by a factor of $dy/dt$.},
In our simulations, however, we found that the error could cause
spurious dissolution of (originally) bound fragments.

We therefore corrected the error as follows.
At every time step, we calculate the expected change in the Jacobi
integral \eqref{Jacobi} due to cooling and to the stresses $G_{xy}+H_{xy}$ acting on
the mean shear. This is accurate to first order in the time step
$\Delta t$. The change in $\Gamma$ computed by ZEUS over the same step
is slightly different.  The error can be predicted to $O(\Delta
t)$ in terms of the state variables at the beginning
of the step and the finite-difference algorithm that updates them.
We compensate for this predicted error by multiplying the internal
energy in every cell by a common factor.\footnote{It might be better
  to use a local correction based on the density gradient and
  velocities, but since the gravitational interactions are intrinsically
  nonlocal, we were unable to find a satisfactory measure of the local error.}
This is equivalent to an extra cooling or heating term.
The required change in the internal energy
is never more than 1\% per time step in the simulations reported here.
Based on simulations of Jeans collapse and other tests,
we believe that this procedure is sufficient to identify the boundary
in parameter space between sheets that fragment and those that do
not.   Unfortunately however, the residual energy errors prevent us
from following the merging of fragments to very large masses.
% In our simulations when cooling and shearing sheet periodic boundary
% conditions are included, bound, isolated fragments will not be
% dissolved due to the numerical effect as can be seen in the examples
% of \S\ref{sec:Results}, which is very important for us because then
% we can trust the code to study the condition of fragmentation. In
% this way, we correct the error globally but not necessary locally.
% And it is true when cooling is strong and numerical error is huge,
% we have to stop the code. However, for our purpose the code is good
% enough for us to study the conditions of fragmentation, which we
% will describe in next section.
Lagrangian methods, such as N-body methods and Smooth Particle
Hydrodynamics, avoid this particular source of error because the non-dissipative
parts of the numerical equations of motion, including gravitational terms, are fundamentally
hamiltonian and have energy integrals (e.g. \citealt{MonaghanPrice2001}).

When fragmentation occurs, the local cooling time $t_c=U(t)/\Lambda$
can become very short in the low-density regions between fragments,
which contain very little mass.  To prevent rapid cooling from
limiting the time step, the internal energy is updated according to
the stable scheme
\begin{equation}
U(t+\Delta t)=\frac{U(t)}{1+\Delta t / t_c}\,, \label{coolingmethod}
\end{equation}
so that that $U$ remains positive for all $\Delta t$.

\section{Results}\label{sec:Results}
In this section, we present results from our simulations.  The
black-hole mass is taken to be $10^8 \msun$, except in
\S\ref{Results:CaseIII} where $\mbh=10^9\msun$.  \S\ref{Results:CaseI} and
\S\ref{Results:CaseII} illlustrate nonfragmenting and fragmenting
regimes, respectively.  In \S\ref{Results:Generalpicture}, we
characterize the boundary between these regimes more systematically.
Except where otherwise noted, all simulations are performed for domain
sizes $L_x=L_y=10 L_0$ at resolution $N_x=N_y=256$.  Experimentation
indicates that varying these numerical parameters upward or downward
by factors $\sim 2$ makes little difference to the results, at least
through the early stages of fragmentation.

We start all simulations with uniform surface density and internal
energy, but with small random velocity perturbations added to the
equilibrium velocity field $\bv=-1.5\Omega x \mathbf{e}_y$.  Apart
from these perturbations and from the numerical parameters cited
above, the initial conditions are therefore characterized by three
physical parameters in addition to the black hole mass: the reference
radius $r_0$, or equivalently the angular velocity
$\Omega=(G\mbh/r_0^3)^{1/2}$; the initial surface density, $\Sigma_i$;
and the initial internal energy per unit area, $U_i$.

\begin{figure}
\centering
\subfigure[]{\includegraphics[width=0.49\hsize]{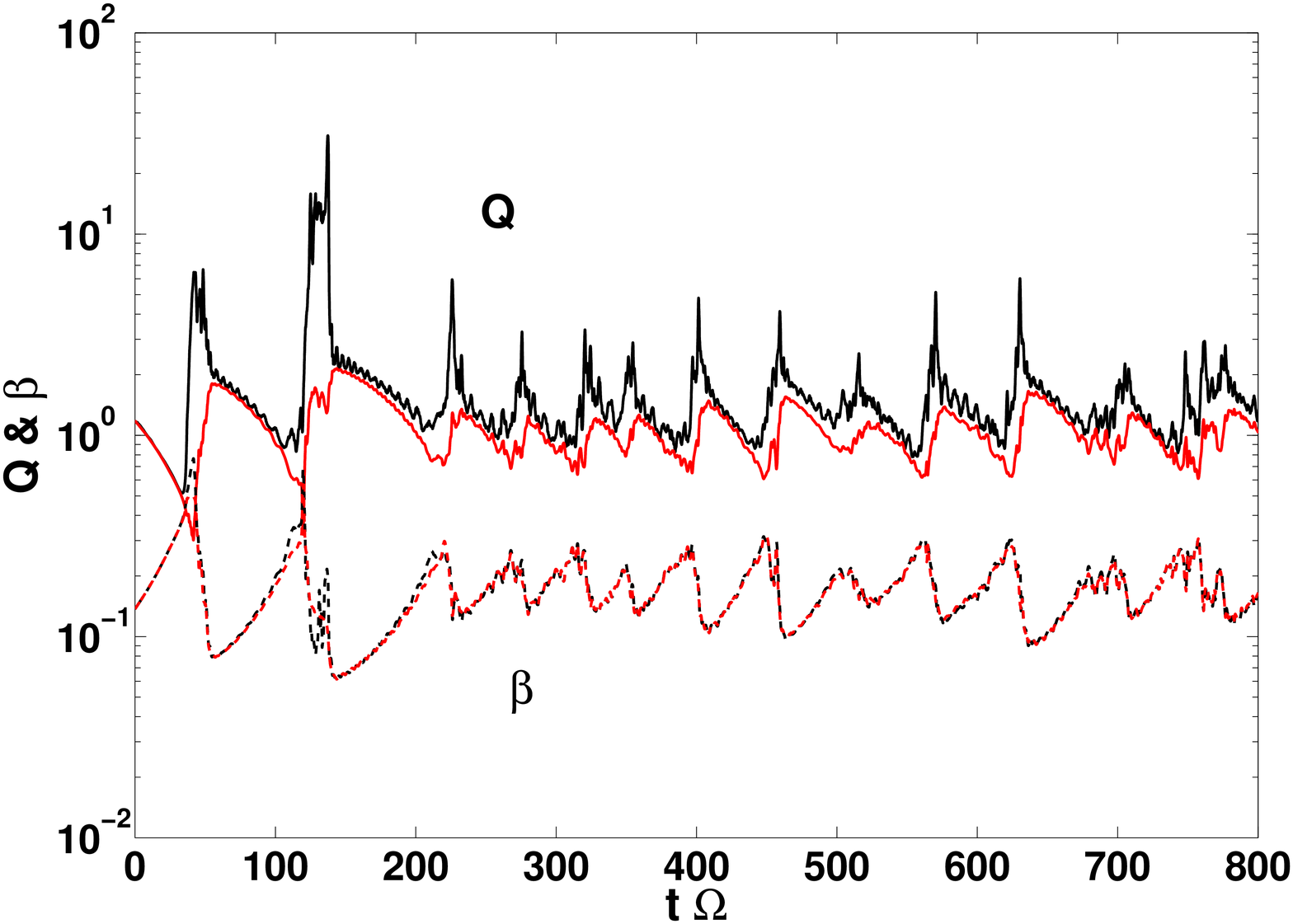}}
\hfill
\subfigure[]{\includegraphics[width=0.49\hsize]{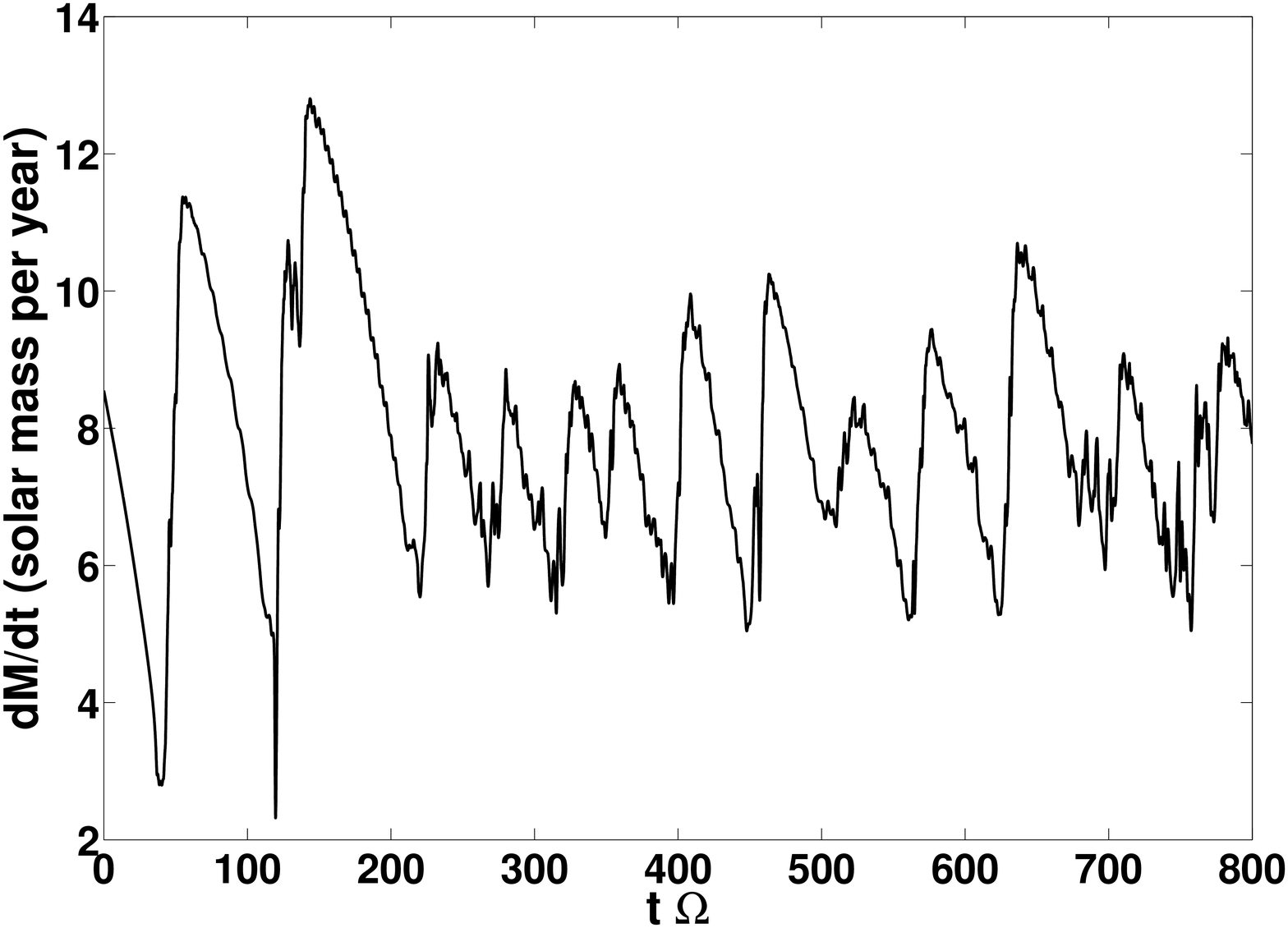}}\\
\vspace{5mm}
\subfigure[]{\includegraphics[width=0.49\hsize]{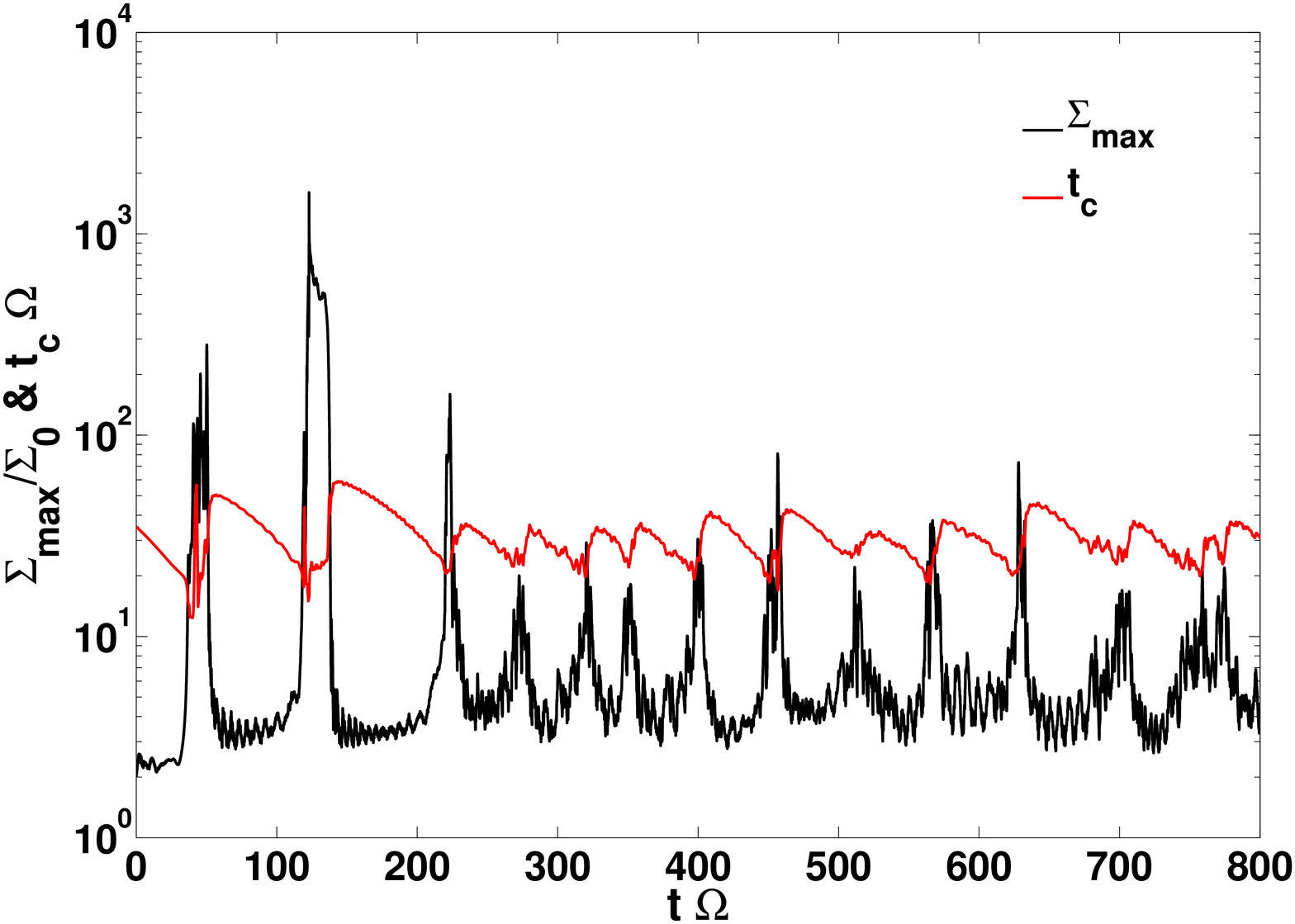}}
\hfill
\subfigure[]{\includegraphics[width=0.49\hsize]{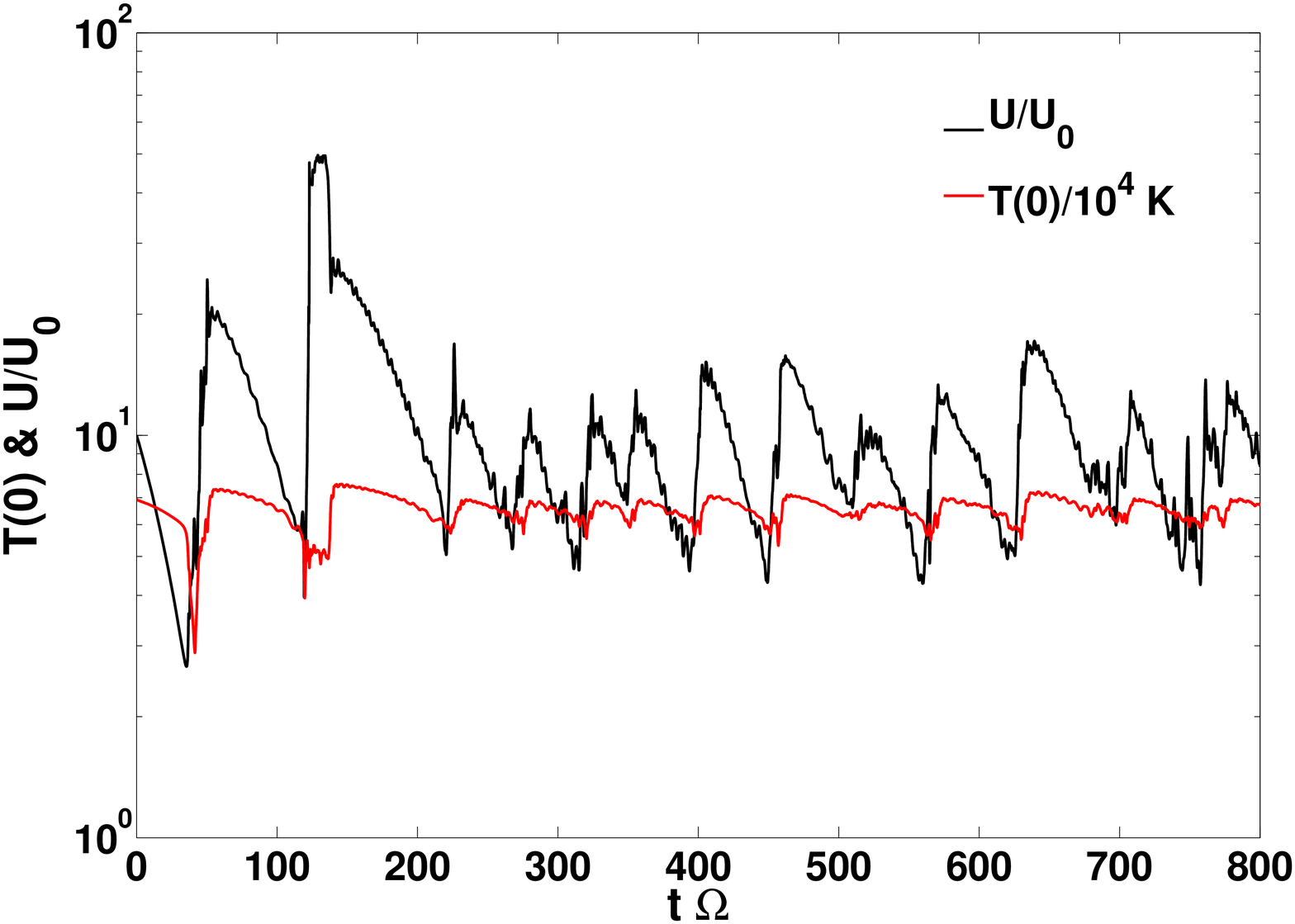} }\\
\vspace{5mm} \caption{Time evolution
of a shearing sheet at mean radius $r_0=10^3 r_s$ about a $10^8 \msun$
black hole, with mean surface density $\langle\Sigma\rangle_A=2\Sigma_0$,
box size $10 L_0\times 10L_0$, and resolution $256^2$. (See
\S\ref{sec:units} for definitions of the units $L_0$ and $\Sigma_0$.)
{\it Panel (a),  solid lines:} Averaged Toomre parameters $\langle Q\rangle_A$ (black);
$\langle Q\rangle_M$ (red). {\it Dashed lines}: averaged
gas-pressure fractions
$\langle\beta\rangle_A$ (black); $\langle\beta\rangle_M$ (red).
{\it Panel (b):} Average accretion rate $\langle\dot M\rangle_A$ via
eq.~\eqref{mdot1} [$\msun{\,\rm yr^{-1}}$].
{\it Panel (c):}  Average cooling time $\langle t_c\rangle_A$ (red) and maximum
surface density $\Sigma_{max}$ (black). {\it Panel (d):}
Average internal energy per unit area $\langle U\rangle_A$ (black)
mid-plane temperature $\langle T(0)\rangle_A$ (red). No
permanent fragments form, so areal and
mass-weighted averages $\langle\ldots\rangle_{A,M}$ are similar. }
 \label{no_fragment_r_1}
\end{figure}

\subsection{Case I: No permanent fragments}\label{Results:CaseI}

Figure~\ref{no_fragment_r_1} shows the evolution of several diagnostic
quantities in a simulation for
$r_0=10^3r_s$, $\Sigma_i=2\Sigma_0$, and $U_i=10U_0$.
The quantities shown in the plots are implicitly averaged over the
grid, and in some cases weighted by mass.
When it is important to be explicit, we use angle brackets with
appropriate subscripts, e.g.
\begin{equation}
  \label{average_notation}
  \langle\beta\rangle_A=\frac{1}{L_x L_y}\iint\beta(x,y,t)\,dxdy\,,
\qquad
  \langle\beta\rangle_M=\frac{1}{L_x L_y\langle\Sigma\rangle_A}\iint\Sigma\beta\,dxdy\,.
\end{equation}
Occasionally overbars are used instead of brackets when we want to
emphasize the time dependence of a spatial average, and we indicate by
context or in some other way whether weighting by mass has been applied.
Double brackets $\avg{\ldots}_A$ or $\avg{\ldots}_M$ will indicate
averages over time as well as space.

In this notation, the areal average
$\bar\Sigma(t)=\langle\Sigma\rangle_A=\Sigma_i$
since mass is conserved by our equations.  
For these choices of $r_0$ and $\bar\Sigma$ mentioned above, we find that
a statistical steady state is reached after $t\sim 200\Omega^{-1}$ in which all of the quantities
shown in Figure~\ref{no_fragment_r_1} fluctuate around long-term averages
$\avg{\ldots}$ that appear to be independent of the initial
value $U_i$, as long as $U_i$ is not so low that the sheet immediately
fragments without cooling.  In particular, $\avg{Q}_A\approx\avg{Q}_M\approx 1$.
Unless otherwise noted, we prefer to start from a hot state $Q_i\gg1$.

To verify the steady state, we compare the cooling and heating rates.
Averaged over the time interval between 200 $\Omega^{-1}$ and 800
$\Omega^{-1}$, the cooling time $\avg{t_c}\approx 29.9\Omega^{-1}$,
and the viscosity parameter $\avg{\alpha}\approx 0.035$, so that
$\avg{t_c}\Omega\approx\avg{\alpha}^{-1}$ as required by thermal
equilibrium (\S\ref{sec:accretion}).  The accretion rate [Panel (b)]
is somewhat larger than the Eddington rate $\dot M_{\rm Edd}=0.22
M_8\epsilon^{-1} \msun{\,\rm yr^{-1}}$ for this black hole for
canonical values of the global radiative efficency
$\epsilon\equiv L_{\rm disk}/\dot M c^2\approx 10\%$.  However, because $r_0$
enters the shearing-sheet equations only via $\Omega$,  these results
could be mapped to $r_0\approx 200\rS$ around a $10^9\msun$ black
hole, where the accretion rate would be sub-Eddington.
Panel (c) shows occasional strong peaks in surface density; however
$\langle Q\rangle$, which is a reciprocal measure of midplane density,
never drops far below unity, showing that no permanent fragments form.
Notice that the mass-weighted average $\langle Q\rangle_M$ is
systematically less than the areal average $\langle Q\rangle_A$
because $\rho(0)$ correlates with $\Sigma$.

\begin{figure}
\centering \subfigure%[surface density $\Sigma$: Red means $\Sigma\ge
%10\Sigma_0$ and black means $\Sigma\approx 0$.]
{\includegraphics[width=0.46\hsize]{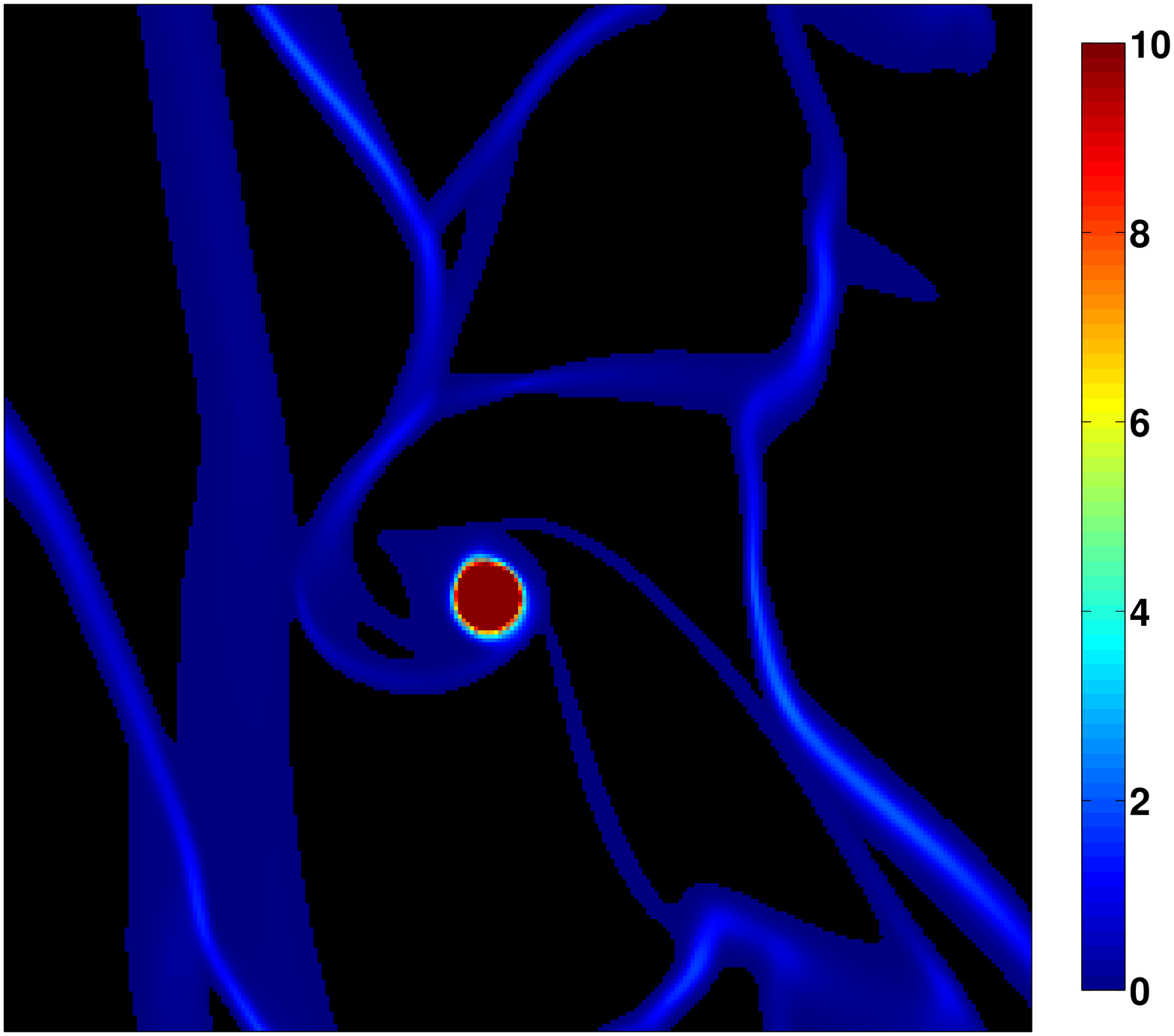} }
\hspace{5mm} \subfigure%[potential vorticity $\xi$ (equation
%\ref{xi}): Red means $\xi\ge 3$ and black means $\xi\le -3$.]
{\includegraphics[width=0.46\hsize]{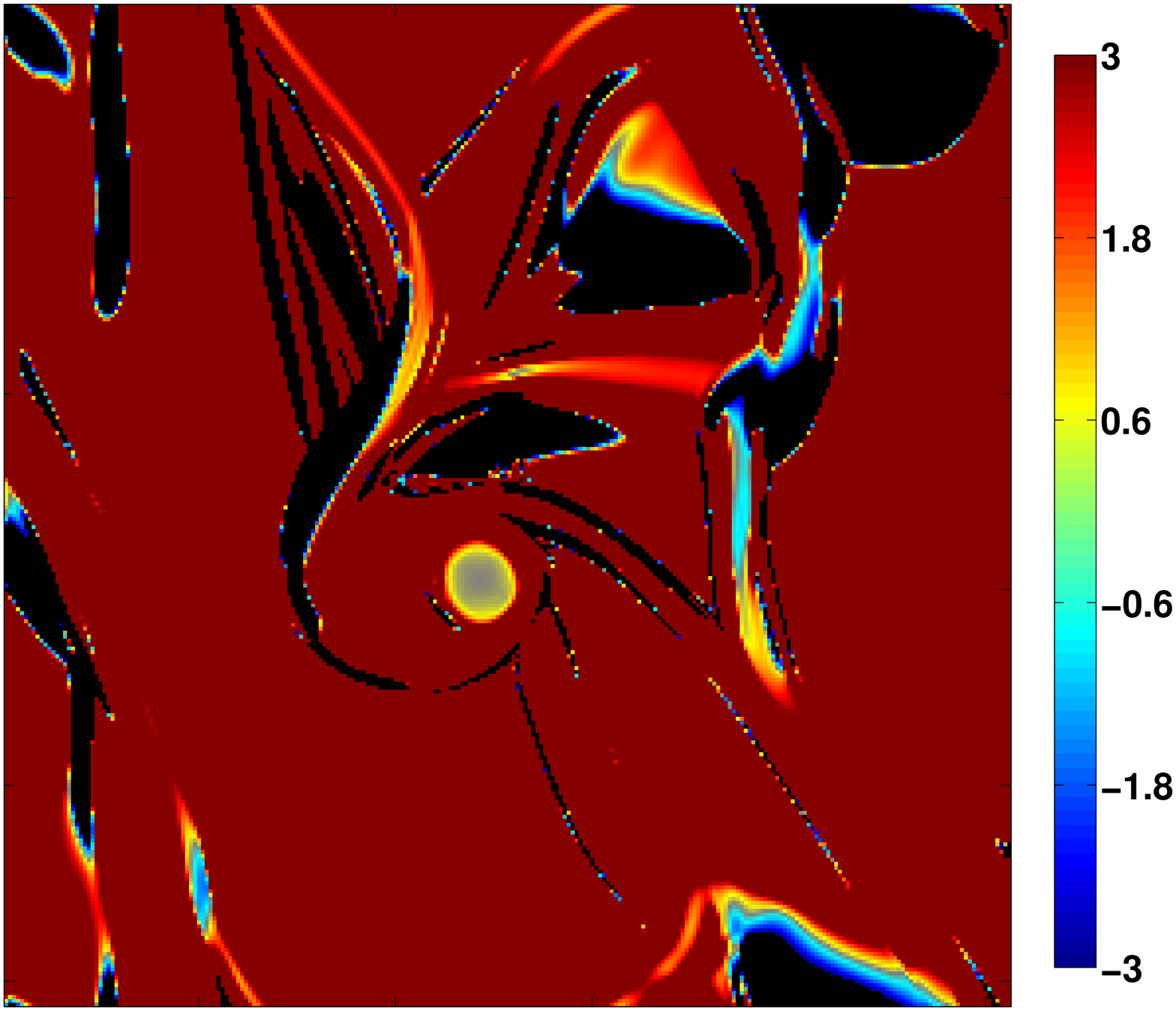} }\\
\subfigure%[$Q$: Black means $Q\ge1$; Red means$Q\approx0$]
{\includegraphics[width=0.46\hsize]{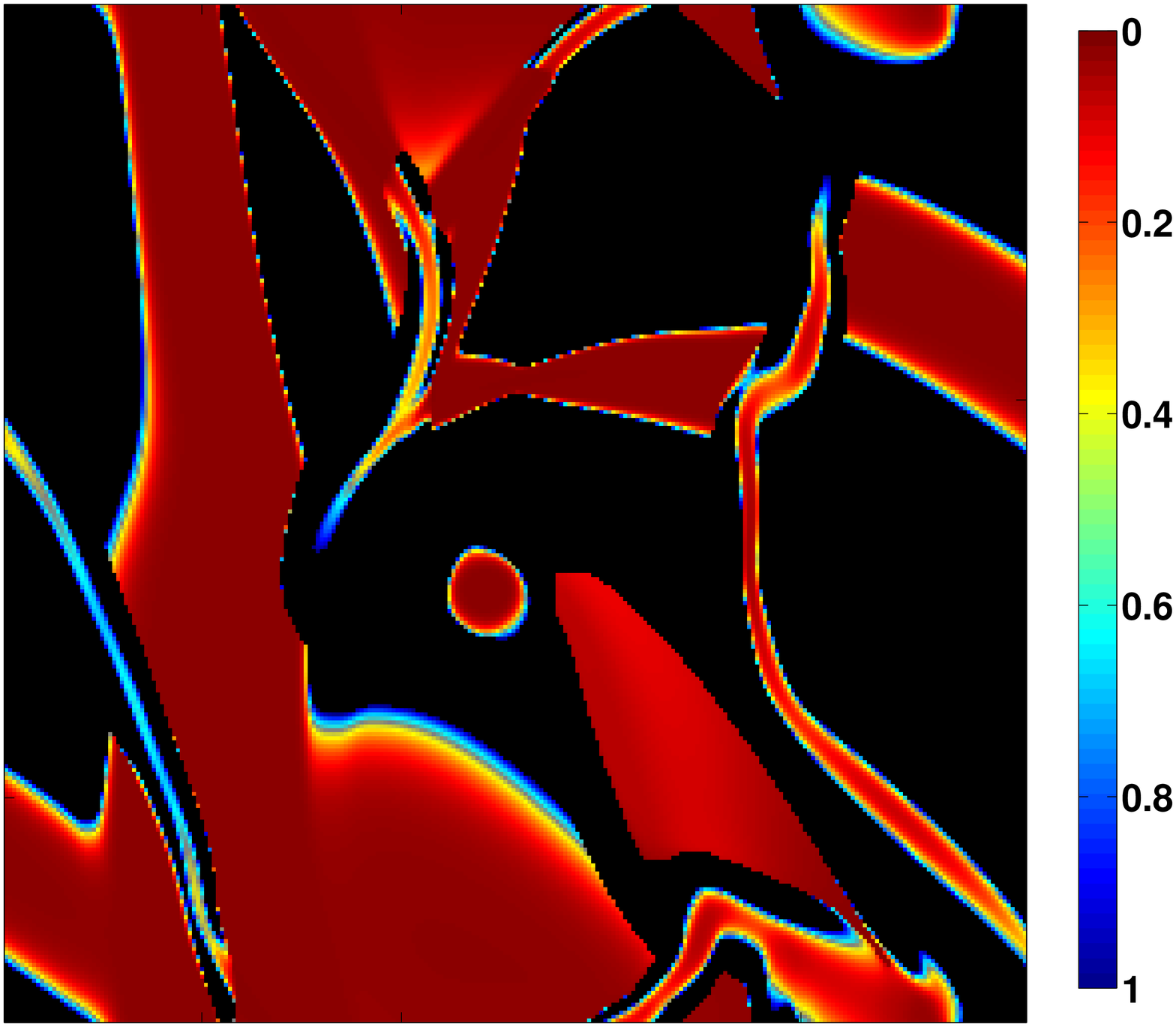}}
\hspace{5mm} \subfigure%[$\beta$: Red means $\beta\approx 0$ and
%black means $\beta\approx 1$.]
{\includegraphics[width=0.46\hsize]{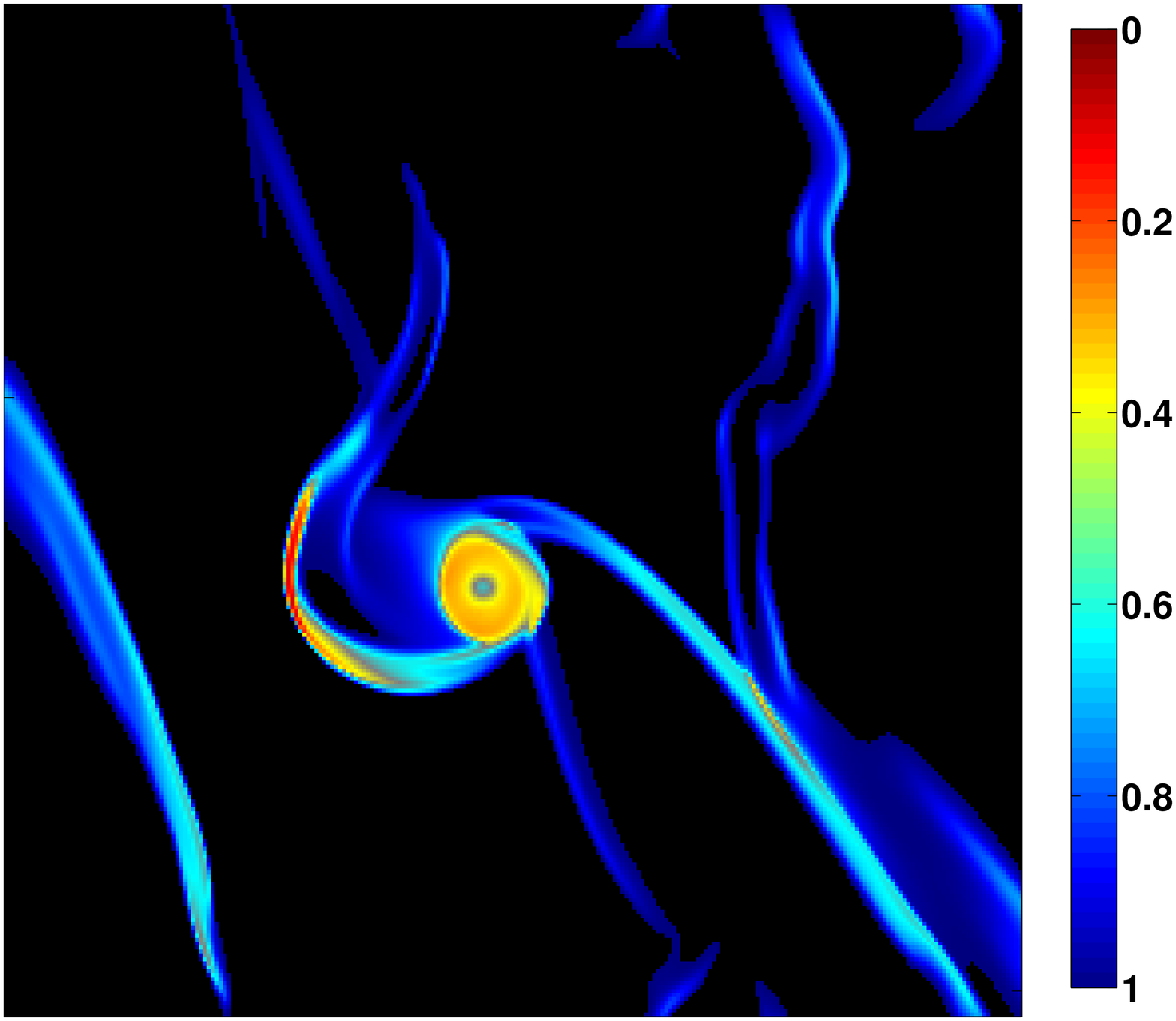}}\\
\caption{A bound fragment at $t=53\Omega^{-1}$ in a simulation for
$r_0=4\times10^3r_s$ and $\langle\Sigma\rangle_A=0.7\Sigma_0$.
{\it Clockwise from upper left:} Surface density $\Sigma/\Sigma_0$;
potential vorticity $\xi$ ($\xi=0.71$ in the initial
uniform state); gas pressure fraction $\beta$; and local Toomre
parameter $Q$ [eq.~\eqref{Q}].}
\label{fragments_S_0_7_r_4}
\end{figure}

\begin{figure}
\centering
\includegraphics[width=0.95\hsize]{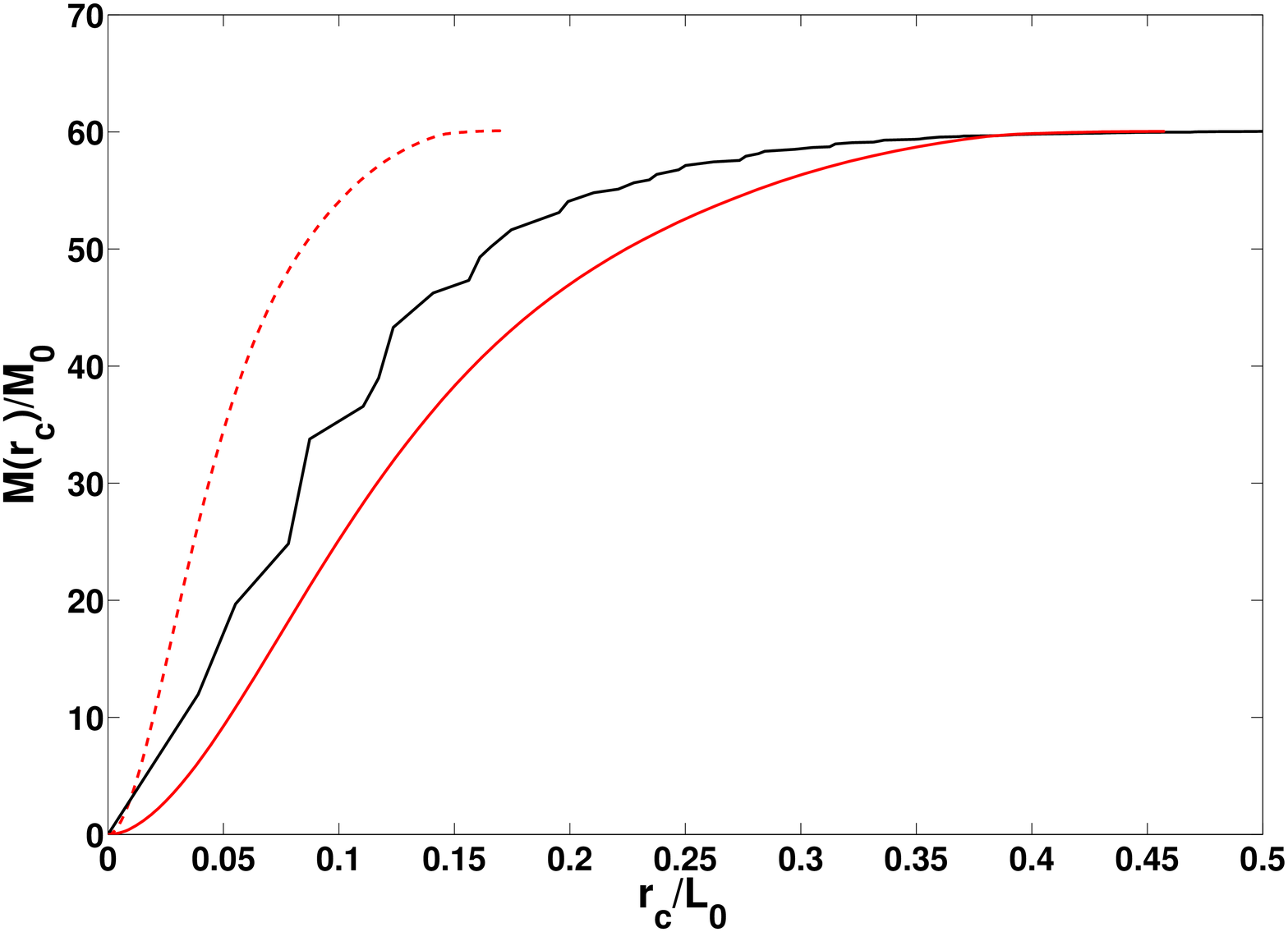}\\
\caption{{\it Black curve:} Density profile of the fragment in Figure
\ref{fragments_S_0_7_r_4} versus distance $r_c$
from its center. Some $99.9\%$ of the bound mass lies at $r_c\le 0.46L_0$,
whereas the Hill radius for $600M_{\odot}$ is $\approx 1.46\,L_0$.
{\it Solid red curve:} Radial profile an Eddington model of
mass $60.1M_0$ and outer radius $0.46L_0$ projected into two
dimensions.  {\it Dashed red curve:} A projected Eddington model of the
same mass having the same central surface density as the fragment.
The object appears to be distended by a combination of rotation and
numerically softened gravity.}
\label{radialprofile}
\end{figure}

% We have explored different surface densities (and thus accretion
% rate) at different radii to study the behavior of the disk. The
% results are summarized in \S\ref{Results:Generalpicture}.
% At each radius, when the surface density is smaller than
% a certain value and accretion rate is small enough, cooling is not
% fast enough and gravitational turbulence is too weak. Then no
% permanent fragment can be formed. For one Eddington accretion rate,
% this is true to $\sim 3\times10^3r_s$. Those simulations confirm
% part of the predictions
% in \cite{Goodman2003} and \cite{GoodmanTan2004} in that quasar disks cannot be
% in a uniform, quiet accretion state in those regimes because
% of the self-gravity of the disk. The fluctuations we see in those solutions are larger
% when the accretion rate is larger.
% However, the unexpected results are that unlike the
% predictions in \cite{GoodmanTan2004}, around one Eddington accretion rate at
% radii from $10^3r_s$ to $3\times10^3r_s$, fragments will not be formed in those regions.
% Then quasar disks can extend to those regions but with large density fluctuations when
% accretion rate is smaller than one Eddington accretion rate.

\subsection{Case II: Permanent fragments} \label{Results:CaseII} 

In a constant-$\dot M$, constant-$\alpha$ disk, $Q$ declines with
increasing radius (e.g. \citealt{Goodman2003}), making fragmentation
more likely.  In a simulation with
$\langle\Sigma\rangle_A=0.7\Sigma_0$ at $r_0=4\times10^3r_s$, the disk
cools to $Q<1$, and fragments form; as the sheet passes through
$\langle Q\rangle_A=1$ the dimensionless cooling time $\Omega\langle
t_c\rangle_A\approx0.5$.
Starting from a uniform state, the mass first concentrates into
azimuthal filaments, which then fragment into several dense
clouds. After merging, a single bound object containing most of the
mass results from our standard $10L_0\times 10L_0$ simulation
(Fig.~\ref{fragments_S_0_7_r_4}). 
Unable to collide with itself, the object steadily cools.  It shows
no tendency to subfragment, as its Kelvin-Helmholtz time (\ref{tKH})
is longer than $\Omega^{-1}$, which in turn is longer than its
internal dynamical time.

No steady state results since we omit fusion reactions (which
would ignite below the resolution of our grid).
For the same $\langle\Sigma\rangle_A/\Sigma_0$, however, fragmention
is avoided at the slightly larger radius $3\times10^3r_s$, where we
measure $\avg{\dot M}\approx 2M_{\odot}{\,\rm yr^{-1}}$, congruent
with an Eddington-limited disk feeding a $10^8\msun$ black hole at
$10\%$ radiative efficiency.

Figure~\ref{fragments_S_0_7_r_4} displays the fragmenting simulation
at $t=53\Omega^{-1}$, after the dominant fragment has coalesced.  As
shown by the lower left panel, $Q\ll 1$ within the fragment, meaning
that its midplane density is well above the Roche value.  About half
of the rest of the sheet is also dense at the midplane, but these
regions have very little mass.  The mass in the bound object is about
$60M_0\approx614M_{\odot}$, $86\%$ of the total.  At this time,
$\avg{Q}_M\approx 0.019$, and $\avg{\beta}_M\approx0.45$. Note that
this $\beta$ is larger than what we expect for a nonrotating Eddington
model of the same mass (eq.~\ref{Mbeta}).  This may in part be a numerical
effect of our finite spatial resolution, which softens the
gravitational force: as the radial density profile
in Fig.~\ref{radialprofile}) shows, the half-mass radius of the object is
approximately two cell widths.   Another cause of the discrepancy
may be the large rotational kinetic energy of this object,
$T/|W|\approx 0.13$.

The energy of this object is partitioned as follows: thermal energy
$E_{\rm th}=936E_0$; kinetic energy (measured with respect to its
center of mass, mainly rotational) $E_k=672E_0$; tidal potential
energy $E_{t}=-0.8E_0$; gravitational self-energy
$E_{p}=-5060E_0$. Thus the total energy of the object in its
center-of-mass frame is negative, implying that the object is bound
and doomed to contract indefinitely.

Our 2D approximation facilitates merging because fragments cannot
avoid one another vertically.  The importance of this can be judged by
examining the epicyclic motions of fragments, if one is willing to
assume that the vertical and horizontal epicyclic amplitudes scale together.
The epicyclic energy per unit mass of a free particle
in the Keplerian shearing sheet is
$\dot{x}^2+(2\dot{y}+3\Omega x)^2$ is a constant.  For an isolated
fragment, the corresponding characteristic quantity is
\begin{equation}
E_{\text{epi}}=\frac{1}{2}M\left[\bar v_{x}^2+(2\bar v_{y}+3\Omega
\bar x)^2\right]. \label{epicycleE}
\end{equation}
Here $M$ is the mass of the fragment, while the overbars mark the
position and velocity of its center-of-mass.
We define the (radial) epicyclic amplitude by
$R_{\text{epi}}\equiv\sqrt{2E_{\text{epi}}/(M\Omega^2)}$.
The importance of the third dimension for collisions can be judged by
comparing $R_{\text{epi}}$ to the physical radius of the object,
$R_*$, or to its Hill radius, $R_H$ [eq.~\eqref{Hillradius}].
We presume that the latter is the more relevant comparison, at least
until $R_*/R_H\lesssim 0.1$, because objects within one another's Hill
sphere undergo a complicated motion in 3D that allows many
opportunities for close passage.

At $t=48.4\Omega^{-1}$
in the above-described simulation for $r_0=4\times10^3r_s$,
there are two fragments, with masses $50.6M_0$ and $2M_0$, so that the
Hill radius associated with their combined masses is $R_H\approx 1.4\,L_0$.
The epicyclic amplitude for the smaller mass is
$R_{\text{epi}}\approx3.14L_0$.   Later in our 2D simulation, these two
fragments merge.  We conclude that the merger might have been delayed
or perhaps even avoided in 3D.

In order to explore merging among more fragments, we have performed a
simulation for the same $r_0$, $\langle\Sigma\rangle_A$, and resolution as in 
Figure~\ref{fragments_S_0_7_r_4} but with four times 
the standard box size, i.e. $L_x=L_y=40L_0$ and $NX=NY=1024$.
The first bound fragments appear at $t\approx13\Omega^{-1}$. 
Along one filament, fourteen small fragments form, and eight merge
in pairs.  It is easily shown the two-body problem decomposes in the
shearing sheet into uncoupled motions of the center-of-mass and
relative coordinates, as in free space.  Therefore, adopting the
approximation that the epicyclic motions of the two components of each
pair are uncorrelated until shortly before they merge,
we add their epicyclic amplitudes in quadrature, $r_{\rm epi}=(r_{\rm epi,1}^2+r_{\rm
  epi,2}^2)^{1/2}$, and compare this to the Hill radius based on their
combined mass, $\RH=[G(m_1+m_2)/3\Omega^2]^{1/3}$, with the results
shown in Table~\ref{tab:epifrag}.
\begin{table}[hbtp]
 \begin{tabular}{ccc}
 $(m_1+m_2)/M_0$ & $m_2/m_1$  & $r_{\rm epi}/\RH$\\ \hline
   2.17                          &  0.61           & 3.10 \\
   3.17                          &  0.83           & 2.40 \\
   1.37                          &  0.99           & 1.15 \\
   1.43                          &  0.96           & 1.68 \\
  \end{tabular}
  \caption{Epicyclic amplitudes of merging pairs.}
  \label{tab:epifrag}
\end{table}
The data in the last column suggest that these encounters might have
proceeded somewhat differently in three dimensions.

\begin{figure}
\centering
\subfigure[]{\includegraphics[width=0.46\hsize]{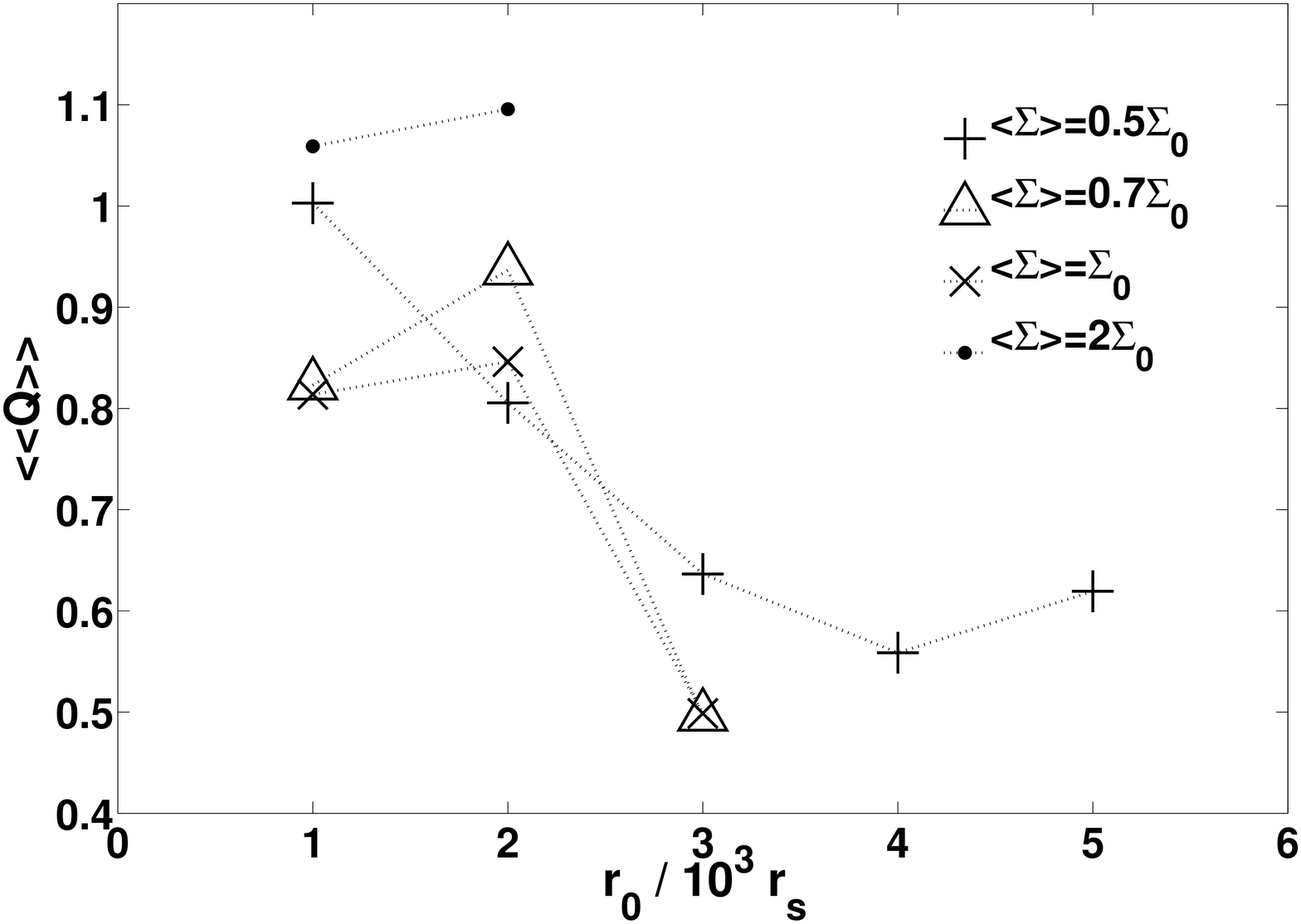} }
\hspace{5mm}
\subfigure[]{\includegraphics[width=0.46\hsize]{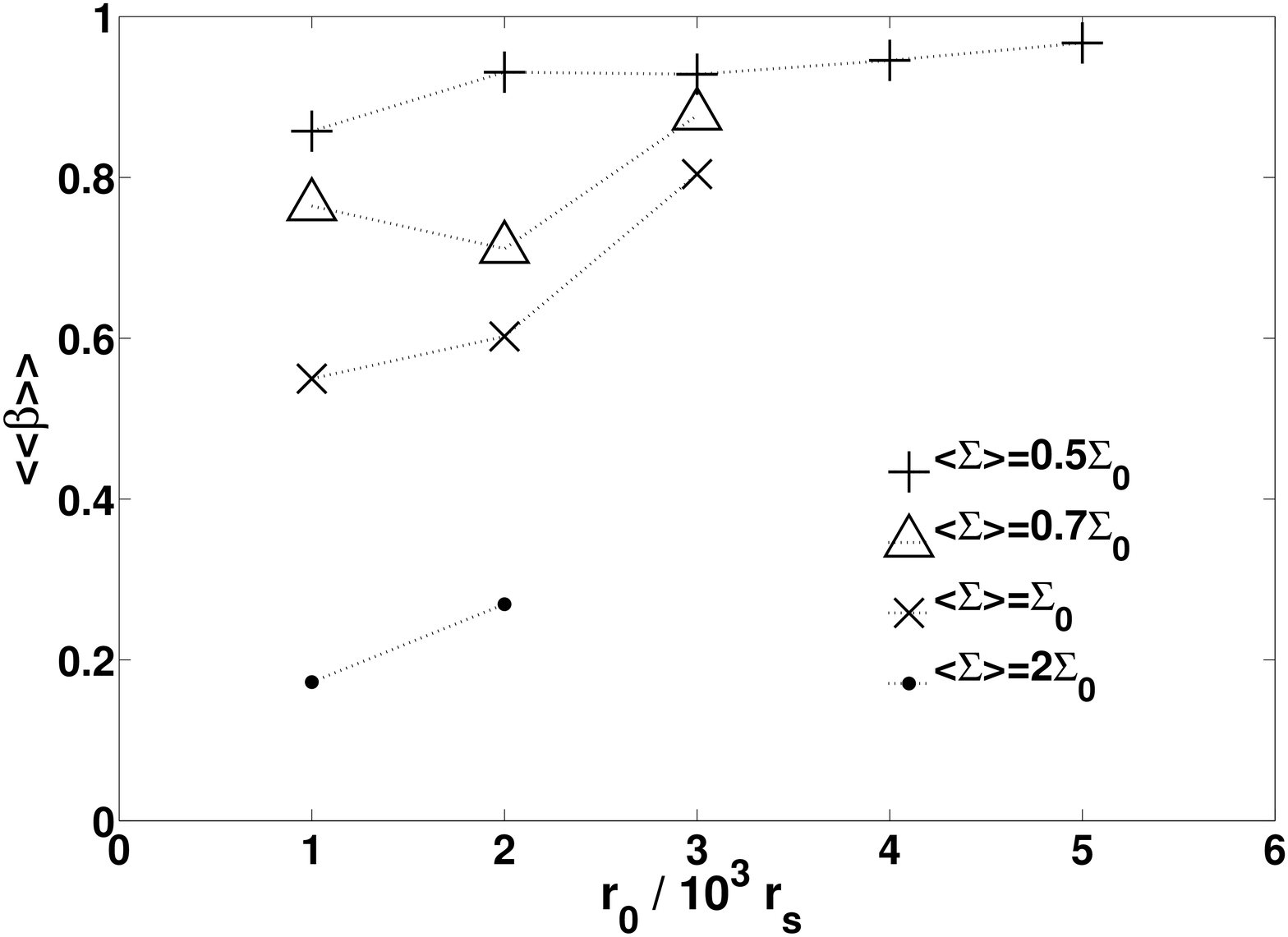} }\\
\subfigure[]{\includegraphics[width=0.46\hsize]{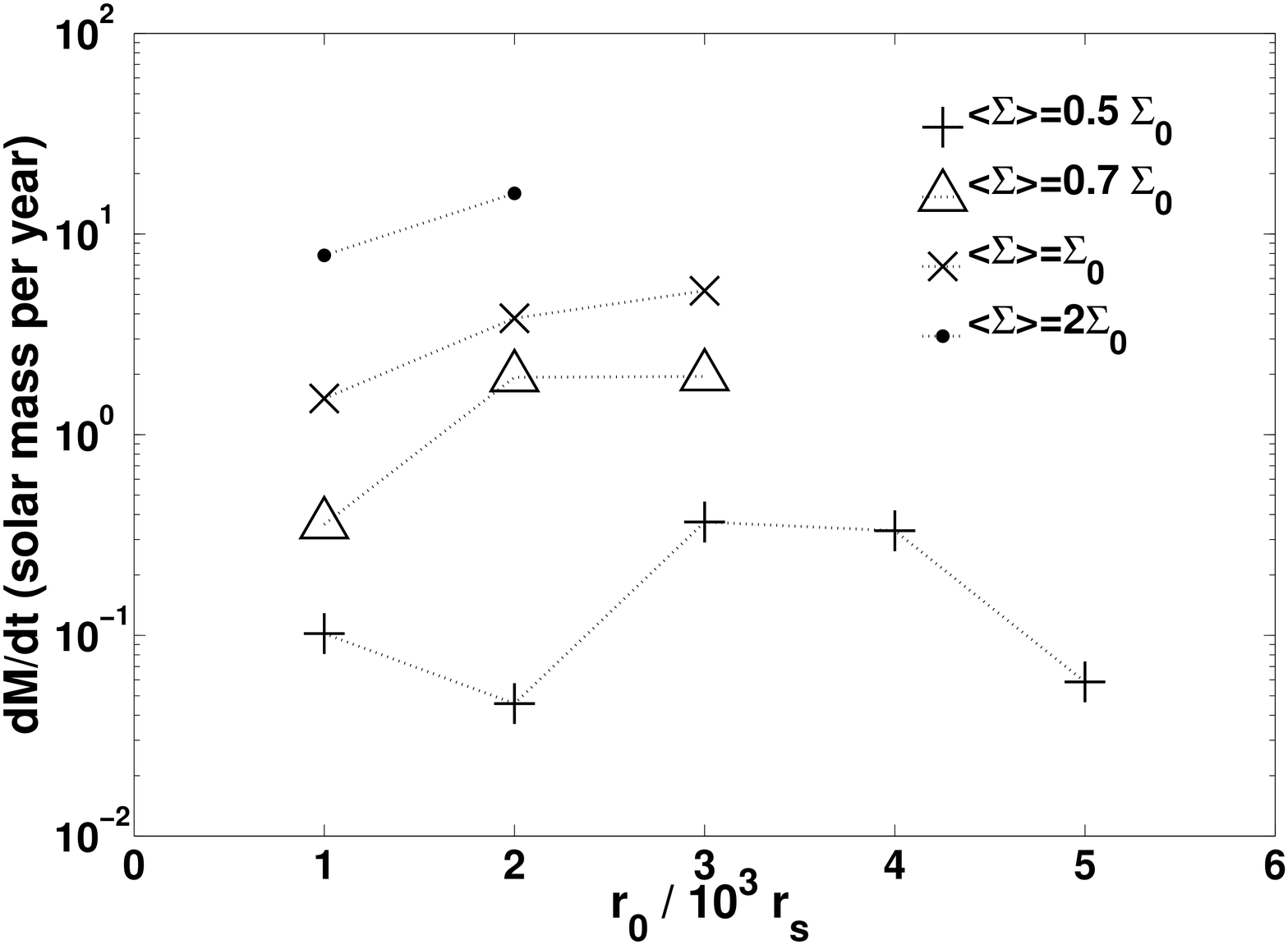}}
\hspace{5mm}
\subfigure[]{\includegraphics[width=0.46\hsize]{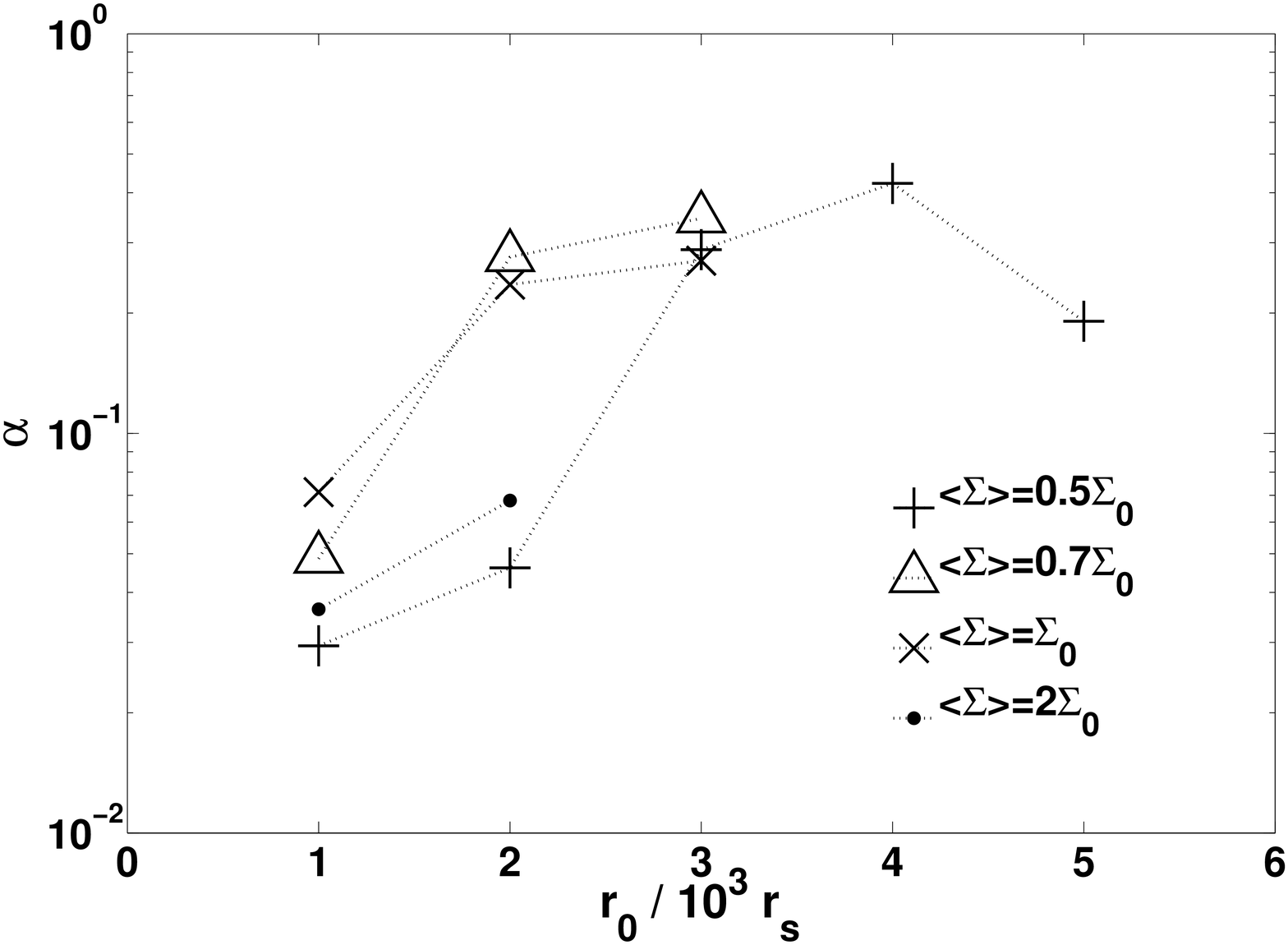}}\\
\caption{Derived properties of statistically steady, nonfragmenting
  simulations versus radius and scaled surface density [$\hat\Sigma$; 
  eqs.~\eqref{Sigma0} \& \eqref{dimlessvars}], for $\mbh=10^8
  \msun$.  The various symbol types mark corresponding simulations in
  all four panels.  The cases for $\hat\Sigma=1,2$ fragment beyond
$2\times10^3\rS$ \& $3\times 10^3\rS$, respectively, hence are not
shown.  The averages of $Q$ and $\beta$ [{\it panels (a)}
    \&{\it (b)}] are mass-weighted and systematically smaller than
  the corresponding areal averages.  
%\remark{By how much? : It depends on 
%how close it is to the boundary of fragmentation. For $\beta$, the 
%difference varies from $10\%$ to $30\%$ of the areal averaged values. 
%For $Q$, it is larger, which varies from $20\%$ to $60\%$ of the areal 
%averaged values.}
}
\label{diskparameters}
\end{figure}

\begin{figure}
\centering
\subfigure[]{\includegraphics[width=0.46\hsize]{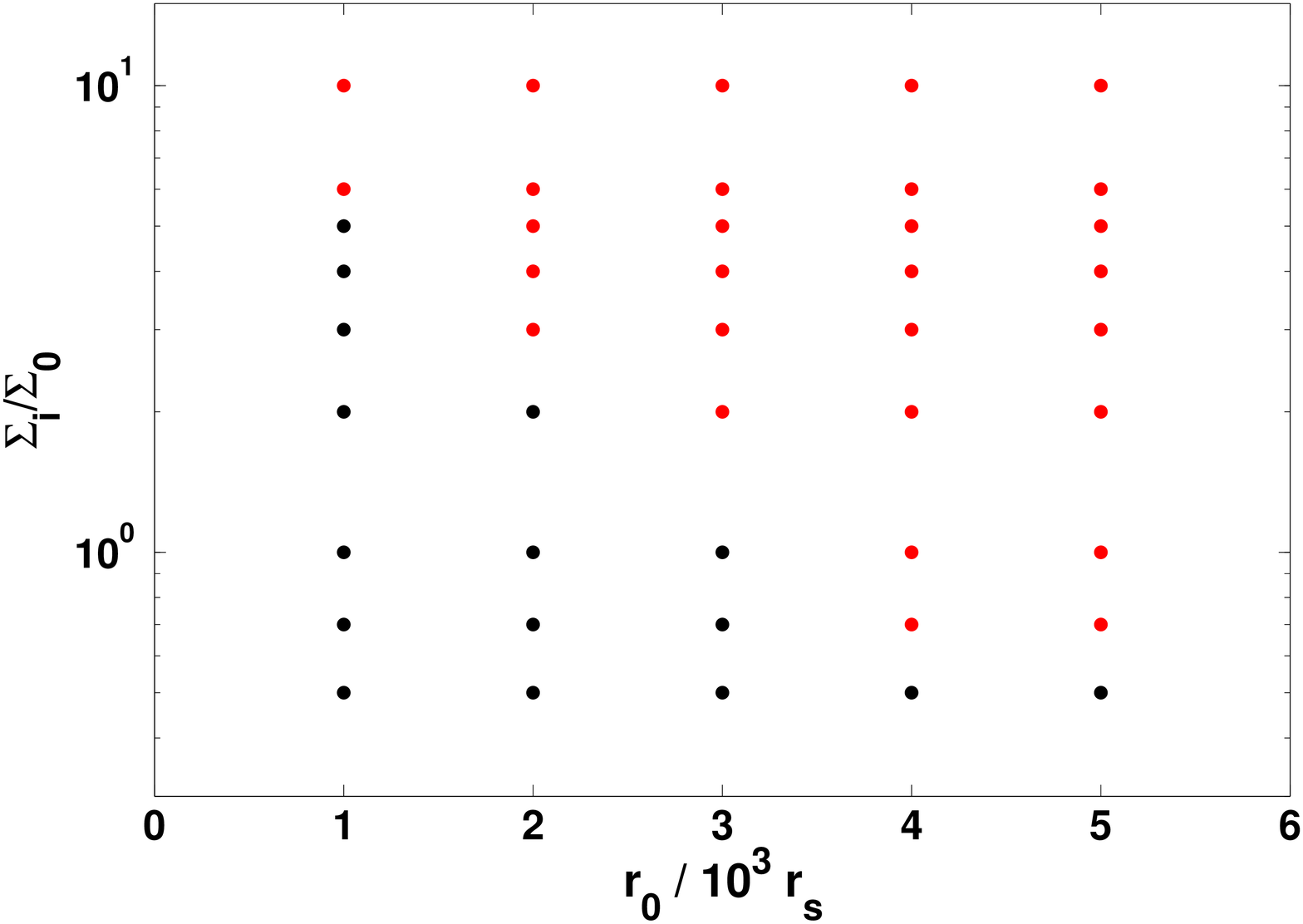}
} \hspace{5mm}
\subfigure[]{\includegraphics[width=0.46\hsize]{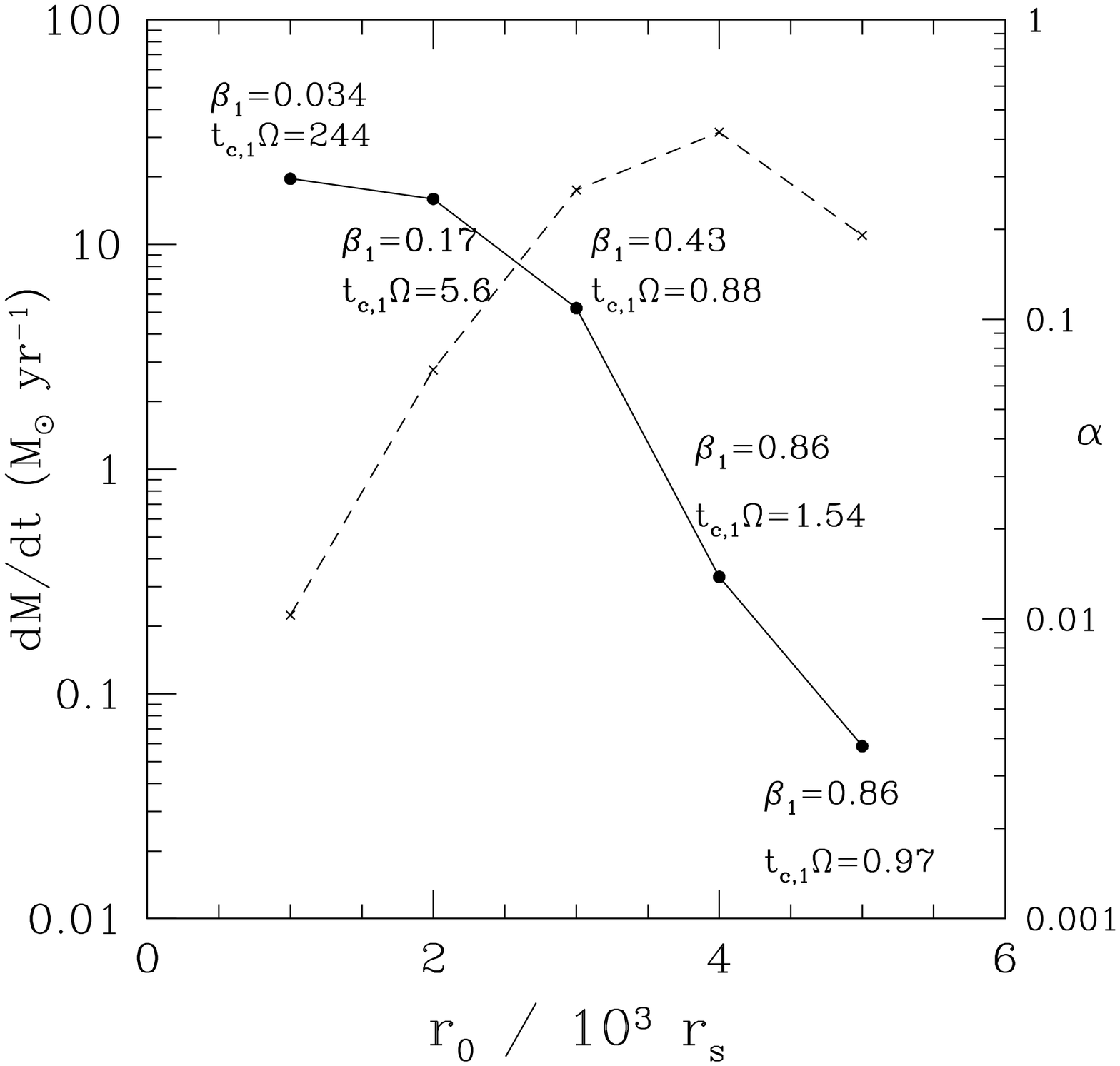} }\\
%\subfigure[]{\includegraphics[width=0.46\hsize]{fragmentboundary_comp} }\\
\caption{The fragmentation boundary. {\it Panel (a):} Red dots mark
  simulations that fragmented, black dots mark those that did not.
    {\it Panel (b):} $\dot M$ (solid line) and $\alpha$ (dashed line) along the
  fragmentation boundary, i.e. for the uppermost black dots in Panel
  (a). These represent maximal rates of gravitational transport
  without fragmentation.  For $\beta_1$ \& $t_{\rm c,1}$, see the
  text.}
\label{fragmentboundary}
\end{figure}

\begin{figure}
\centering
\includegraphics[width=0.46\hsize]{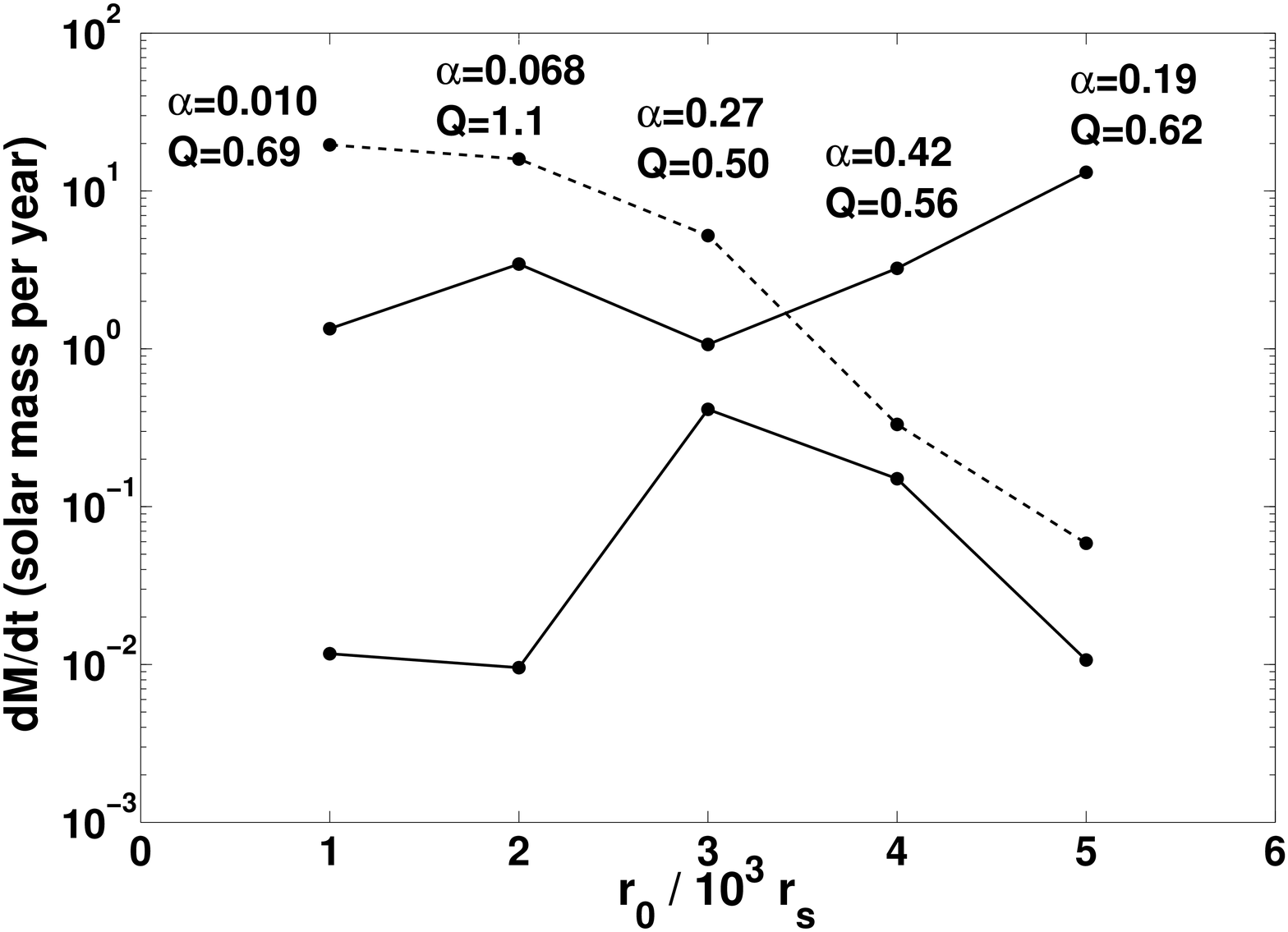}\\
\caption{Comparison of $\dot M$ along the fragmentation boundary with
  G03.  {\it Dashed curve:} As in Panel (b) of
  Fig.~\eqref{fragmentboundary}.  {\it Solid curves:} Predicted values
  of $\dot M$ for the indicated $Q$ \& $\alpha$ taken from the
  simulations, using eqs.~(11) \& (A3) of G03.}
\label{fb_comp}
\end{figure}

\subsection{General picture of the self-gravitating regime}
\label{Results:Generalpicture} In this section, we summarize the
general trends found in our shearing-sheet models, with particular
attention to the conditions for fragmentation.

As discussed above, the eventual statistical steady state of our
simulations, when it exists, are defined by two control parameters,
mean surface density, $\langle\Sigma\rangle$, and angular velocity,
$\Omega=(G\mbh/r_0^3)^{1/2}$.
We have explored the parameter ranges
$0.5\Sigma_0\le \langle\Sigma\rangle\le 10\Sigma_0$ and
$10^3\rS\le r_0\le 5\times 10^3$, with $\mbh=10^8 M_{\odot}$.
Throughout most of this regime, the effective Shakura-Sunyaev
parameter is measured to be $\alpha\gtrsim 10^{-2}$, so that
self-gravity would likely have dominated the angular momentum
transport even had we included MHD in our simulations.
The corresponding accretion rates are
$0.01\lesssim \dot M\lesssim 20\ \solaryr$.  

Figure~\ref{diskparameters} shows several steady-state quantities as
functions of our two control parameters.  With the physics in our
models, steady states are not possible after fragmentation, so these
quantities are measured from simulations that did not fragment.  As
seen in the first panel, the mass-weighted average of $Q$ is typically
slightly less than $\lesssim1$.  Recall that our definition of $Q$
[eq.~\eqref{Q}] is simply a reciprocal measure of the midplane density
relative to the Roche density; under adiabatic compression by a
nonlinear density wave, the internal energy of the gas rises in step
with its density, so that a bound fragment may be avoided even if $Q$
falls briefly below unity.  Mass weighting tends to emphasize these
transiently compressed regions.  The \emph{areal} average of $Q$ is
typically $\gtrsim1$. At a given
$\hat\Sigma=\langle\Sigma/\Sigma_0\rangle$, $Q$ decreases toward
larger radii, where cooling is stronger.  Panel (b) shows that the
mass-weighted $\beta$ increases with increasing radius and decreasing
surface density.  Panel (c) shows that $\dot{M}$ is much more
sensitive to surface density ($\hat\Sigma$) than to radius or,
equivalently, to $\Omega$.  For comparison, the Eddington rate for a
black hole of $10^8 M_8\,\msun$ is $\dot M_{\rm Edd}\approx 2 M_8
\epsilon_{0.1}\solaryr$, where $\epsilon=0.1\epsilon_{0.1}\equiv
L_{\rm disk}/\dot Mc^2$ is the global radiative efficiency of the
disk.  For $\hat\Sigma=1$, the Eddington rate is achieved at
$r_0\approx 3\times10^3r_s$.  At the same $\dot M$, however, our
models fragment when $r_0\gtrsim 4\times10^3r_s$.  The final panel
shows that generally $\alpha$ increases with increasing radius at fixed
$\hat\Sigma$.
% in accord with decreasing $Q$. 
But $\alpha$ has a
complicated dependence on surface density. The largest value
encountered in any of our non-fragmenting simulations was
$\alpha_{\max}\approx 0.4$.

Figure~\ref{fragmentboundary} shows the boundary between those cases
that fragment and those that do not in the plane of
$(r_0,\hat\Sigma)$, our two control parameters.  Each dot represents a
simulation, all done with $L_x=L_y=10L_0$ and $NX=NY=256$. We have
checked that the boundary between the fragmenting (red) and
nonfragmenting (black) cases is not significantly altered at higher
resolution ($NX=NY=512$).
% Most of the results are the same, except at radius $4\times10^3r_s$ and
% $5\times10^3r_s$, where the boundary between fragmentation and no
% fragmentation moves down a little bit. In the high resolution at
% $4\times10^3r_s$, we have to go to $\Sigma_i=0.48\Sigma_0$ so that no
% permanent fragment is formed. At $5\times10^3r_s$, when we go to
% $\Sigma_i=0.5\Sigma_0$, no permanent fragment is formed. This shows
% that the boundary we find here is not sensitive to the resolution.
Higher resolution does make a difference, however, when the most
unstable wavelength is short, which happens when the accretion rate
(and thus surface density) is very small: our standard resolution
begins to fail at $\dot M<0.1\solaryr$.

Panel (a) of Figure \ref{fragmentboundary} shows that the maximum
dimensionless surface density that the disk can support without
fragmenting declines rapidly with increasing radius.  Over the range
$10^3\le r_0/\rS\le 4\times 10^3$, the boundary can be fit to a power
law: $\hat\Sigma_{\rm frag}\approx 6 (r/10^3\rS)^{-1.5}$.  Since, as
shown in Figure~\ref{diskparameters}, the local accretion rate is much
more sensitive to $\hat\Sigma$ than to $r_0$, it follows that $\dot
M_{\rm max}$ also declines swiftly with radius; this is confirmed by
the second panel.  In fact, $\dot M_{\max}$ declines some two orders
of magnitude between $10^3\rS$ and $5\times 10^3\rS$.

The gas pressure fraction $\beta_1$ and cooling time $t_{\rm c,1}$
marked in the second panel
are neither mass-weighted  nor areal averages: instead, they are computed for the
same $r_0$ and $\langle\hat\Sigma\rangle$ as those in the simulation, but
with $Q$ set to unity
in the equation of state rather than its measured steady-state value.
This allows us to compare the observed fragmentation boundary
with Gammie's criterion $(\Omega t_{\rm c})_{\rm crit}=\mbox{constant}=O(1)$.
Except at the innermost radius shown,  we find that the cooling time
is indeed $O(\Omega^{-1})$ along the boundary.
However, fragmentation occurs at $r=10^3\rS$ when
$\langle\hat\Sigma\rangle\gtrsim 5$ even though $\Omega t_{\rm c}\gg
1$; this is possible because $\beta\ll1$, so that bound fragments are
only marginally stable against collapse even without energy loss.
For $\mbh=10^8 \msun$, this regime is reached only at local
accretion rates far above the Eddington rate, but not so for larger
$\mbh$, as will be shown in \S\ref{Results:CaseIII}.

We have compared our simulations with the alpha-disk models of
G03 for the case that viscosity is proportional to
total pressure.  Figure~1 of G03 displays curves of
constant $\alpha$ and $Q$ in a plane of $\dot M$ versus $r$.  For
$Q=1$ and plausible $\alpha$, there are generally two branches to the
curve: a high-$\dot M$ solution, which has high surface density and
low $\beta$, and a low-$\dot M$ solution, which has the oppositie
properties.  These branches join at $r\approx 10^3\rS$, so that $Q>1$
for all solutions at smaller radii.  To compare with these
predictions, we take the measured values of $\alpha$ and mass-weighted
$Q$ from the simulations along the fragmentation boundary shown in
Panel~(b) of Fig.~\ref{fragmentboundary}, and we insert these values
into the model for $\dot M$ from G03.  The results are
shown by solid lines in 
Fig.~\ref{fb_comp}. There are again two solutions for $\dot
M$ at each radius, with radiation pressure dominating the upper
(higher $\dot M$) solution, and gas pressure dominating the lower.  But while
$Q$ is roughly constant along these curves, $\alpha$ is not---$\alpha$
decreases rapidly with decreasing radius.  The actual
$\dot M$ directly measured in the simulations (dashed curve) lies
\emph{above} the upper branch at $r< 4\times 10^3\rS$ and has slightly
higher surface density.  The
differences between the predicted and measured values of $\dot{M}$ may
be due in part to the assumption of uniform conditions in the $\alpha$
models, so that mass and areal averages differ.

\subsubsection{Scaling to other black-hole masses}
\label{sec:scaling}

Apart from fundamental constants and numerical parameters (grid
resolution, domain size, etc.), the statistical steady states of our self-gravitating
shearing sheets are entirely determined by just two control
parameters:\footnote{Actually, the metallicity of the gas
should be counted as a third parameter.  
Since it enters the opacity \eqref{opacity} as well as
our mass unit \eqref{massunit}, it cannot be entirely scaled out of
the simulations, for which  
we have taken $\mu=\mu_\odot$ throughout.}
 $\Omega$ and $\langle\Sigma\rangle$.  Therefore, although
we have fixed $\mbh=10^8 \msun$ in our simulations and studied the
outcomes as functions of $r_0$ and $\langle\hat\Sigma\rangle$, we can
scale our results to other black-hole masses by recasting them in
terms of the two control parameters above.  For ease of writing, we
will omit the angle brackets from $\langle\Sigma\rangle$ and
$\langle\hat\Sigma\rangle$ henceforth.

Our most important result is the fragmentation boundary.
As noted above, $\hat\Sigma_{\rm frag}\approx
6r_3^{-x}$, with $x\approx 1.5$.  Since
$\rS\propto\mbh$ and $\Sigma_0\propto\Omega^{4/3}$
[eq.~\eqref{Sigma0}], this can be recast as
\begin{equation}
  \label{eq:Sigmacrit}
  \Sigma_{\rm frag}\approx 2\times 10^6\left(M_8^{2/3}r_3\right)^{-(2+x)}\,{\rm g\,cm^{-2}}\,.
\end{equation}

We compare this with the surface density required for accretion at the
Eddington rate.  From Panel~(c) of Fig.~\ref{diskparameters}, it
appears that $\dot M$ is much more sensitive to $\hat\Sigma$ than to
radius.  This implies that $\Sigma\propto r^{-2}$ at fixed $\dot M$.  We will attempt to
explain this scaling below, but for the moment, we simply accept it.
Since $\dot M$ increases by a factor $\approx 10^2$ as $\hat\Sigma$
increases by $4$, we estimate $\dot M\propto\hat\Sigma^y$ with
$y\approx 3.3$.  Figure~\ref{diskparameters} also indicates that
$\hat\Sigma\approx 0.7$ yields $\dot M\approx 2{\rm\,\msun
  \,yr^{-1}}$, which is the Eddington rate for
$\mbh=10^8{\rm\,\msun}$ and radiative efficiency $\epsilon=0.1$.
This coincides with $\Sigma_{\rm frag}$ at $r_3\approx 4$ for $M_8=1$.
Rewriting the relation $\dot M\approx 2{\rm\,\msun\,yr^{-1}}(\hat\Sigma/0.7)^y$
as $\dot M\propto(\Sigma\Omega^{4/3})^y$, we find that the radius
beyond which a self-gravitating accretion disk will fragment if it
accretes at a fraction $\dot m$ of the Eddington rate is, taking
$x=1.5$ and $y=3.3$,
\begin{align}
  \label{eq:rcrit}
  r_{\rm crit} &\approx 4\times 10^3 M_8^{-0.87}\dot m^{0.2}\rS
\quad\approx 0.04\, M_8^{0.13}\dot m^{0.2}\,{\rm pc}.
\end{align}
For comparision, \cite{Goodman2003}'s equation (10) predicts that
$Q=1$ in an alpha disk at $r_3\approx 2.7
(\alpha/M_8)^{2/9}$ if the viscous stress is proportional to total
pressure and $\beta\ll 1$.  This is roughly half of
eq.~\eqref{eq:rcrit} for $M_8=1$ and $\alpha=0.4$ (the
largest value found in our simulations), but
the scaling with black-hole mass is different.
As Panel~(b) of Figure~\ref{diskparameters} shows, however, $\beta$ is
closer to 1 than to 0 at $r_{\rm crit}$ for $M_8=1$, so precise
agreement is not to be expected.

We promised to discuss why $\Sigma\propto r^{-2}$ at fixed $\dot M$
and $\mbh$.  When self-gravity controls the accretion rate, $Q\approx
1$, so that the midplane density $\rho(0)\propto r^{-3}$.  It follows
from vertical radiative and hydrostatic equilibrium that the half
thickness of the disk is
\begin{equation}\label{eq:hval}
  h\approx \frac{3\kappa\dot M}{8\pi c} (1+2Q^{-1})^{-1}(1-\beta)^{-1}
\end{equation}
to the extent that $\beta$ is vertically constant.  This gives the
familiar result that $h\approx\mbox{constant}$ in steady disks where
radiation pressure dominates.  Then we would have
$\Sigma=2h\rho(0)\propto r^{-3}$, not $r^{-2}$, for constant $Q$.
Fig.~\ref{diskparameters} shows, however, that $\beta\gtrsim 0.7$ at
$r_3\ge1$ for $\hat\Sigma=0.7$, the value that gives a roughly
Eddington accretion rate for $M_8=1$.   Thus $h$ may vary with radius
through the factor $(1-\beta)^{-1}$.  Now eq.~(A3) of
\cite{Goodman2003} predicts that
\begin{equation}
  \label{eq:1minusbeta}
  (1-\beta)^{-1}\beta^{1/2+(b-1)/10}\approx 0.35(\alpha_{0.1}M_8)^{-1/10}\dot m^{-4/5}
\left(\frac{\kappa}{\kappa_{\rm es}}\right)^{-9/10} r_3^{21/20}\,,
\end{equation}
where $b=0$ or $b=1$ according as $\nu\propto P$ or $\nu\propto P_{\rm
  rad}$: clearly it makes little difference to the value of
$(1-\beta)$ when this is $\lesssim 0.5$.  Although $\alpha$ is not constant with radius in our
self-gravitating models, the dependence in eq.~\eqref{eq:1minusbeta}
is so weak that $(1-\beta)^{-1}$ and hence $h$ are approximately
linear in $r$ when $\beta\gtrsim 0.5$.  This explains why
$\Sigma\propto r^{-2}$, but it also shows that this scaling holds only
over a limited range of $r$ and $\mbh$.  

Equation \eqref{eq:rcrit} shows that self-gravity is important at a
smaller multiple of $\rS$ for larger $\mbh$ at a given Eddington
fraction $\dot m$; it then follows from eq.~\eqref{eq:1minusbeta} that
the self-gravitating regime is characterized by smaller $\beta$ for
larger $\mbh$.  In fact, for $\mbh\gtrsim10^9{\rm\,\msun}$, we
estimate that $\beta<0.1$ at $r_{\rm crit}$, so that fragmentation may
occur with little cooling, as demonstrated in \S\ref{Results:CaseIII}.
Thus, while it remains true that the local dynamics of a
self-gravitating disk is determined by $\Omega$ and $\Sigma$, the
particular scaling \eqref{eq:rcrit}, which depends upon our power-law
fit to the fragmentation boundary over a limited range of $\beta$, is
likely to be modified for black-hole masses much above
$10^8{\rm\,\msun}$.  On the other hand, for black holes much less
massive than our fiducial value, Kramer's opacity will dominate over
electron scattering at $r_{\rm crit}$; in view of the sensitivity of
the radiation fraction \eqref{eq:1minusbeta} to $\kappa$, this also
will modify eq.~\eqref{eq:rcrit}.  Thus, the latter equation is
probably quantitatively reliable within only a narrow range around
$M_8=1$.  Nevertheless, the trend is surely correct: namely, that
$r_{\rm crit}$, the radius beyond which accretion at the Eddington
rate would cause fragmentation, occurs at a smaller multiple of $\rS$
for larger $\mbh$.

\begin{figure}
\centering \subfigure%[Evolution history]%[Surface density $\Sigma$: Red means
%$\Sigma\ge20\Sigma_0$ and black means $\Sigma\approx0$
%($t=265\Omega^{-1}$).]
{\includegraphics[width=0.7\hsize]{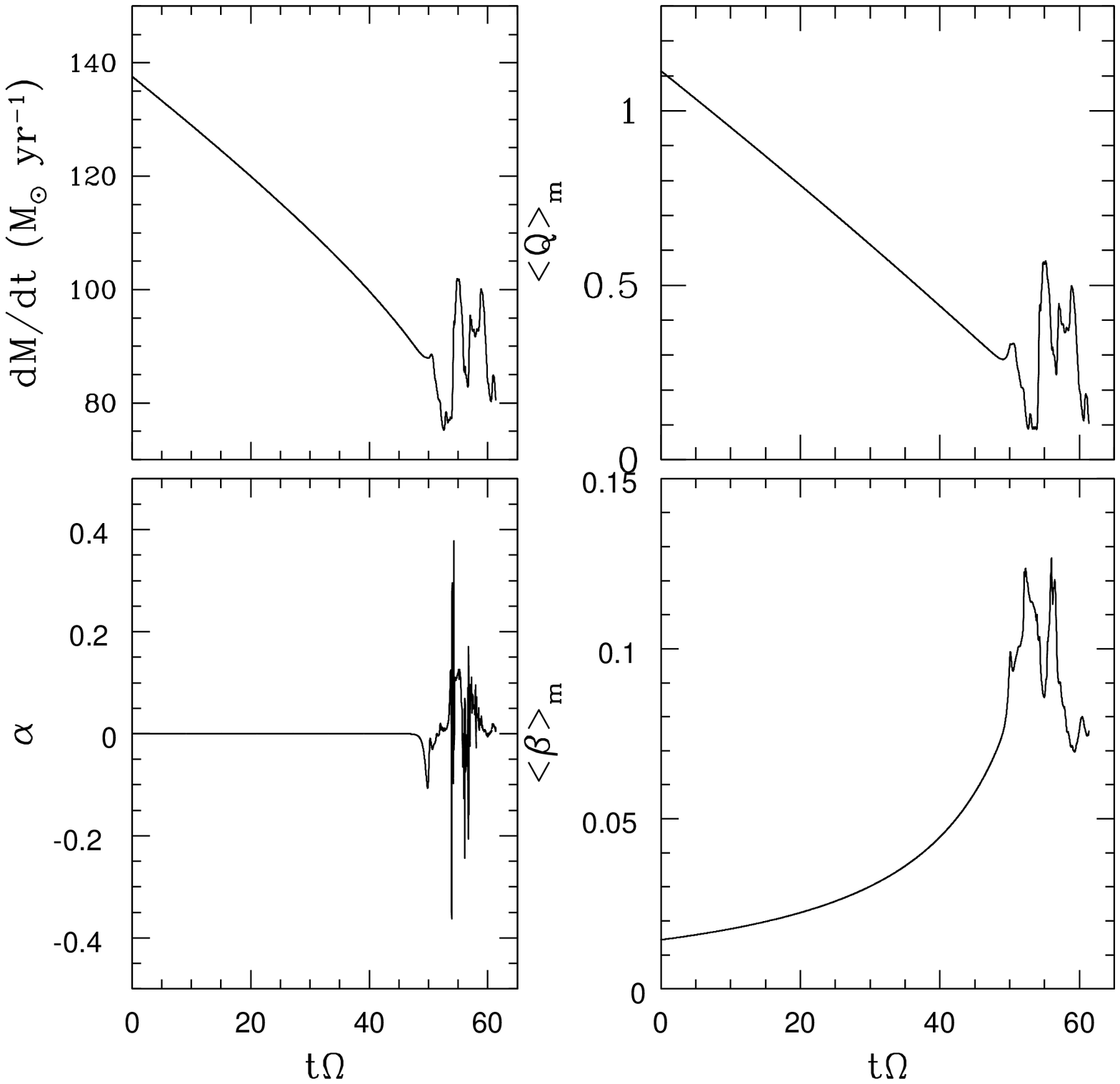} }\\
\vspace*{-0.3in}
\subfigure[$t\Omega=53$]%[Radial profile for the upper fragment in the left plot.]
{\includegraphics[width=0.4\hsize]{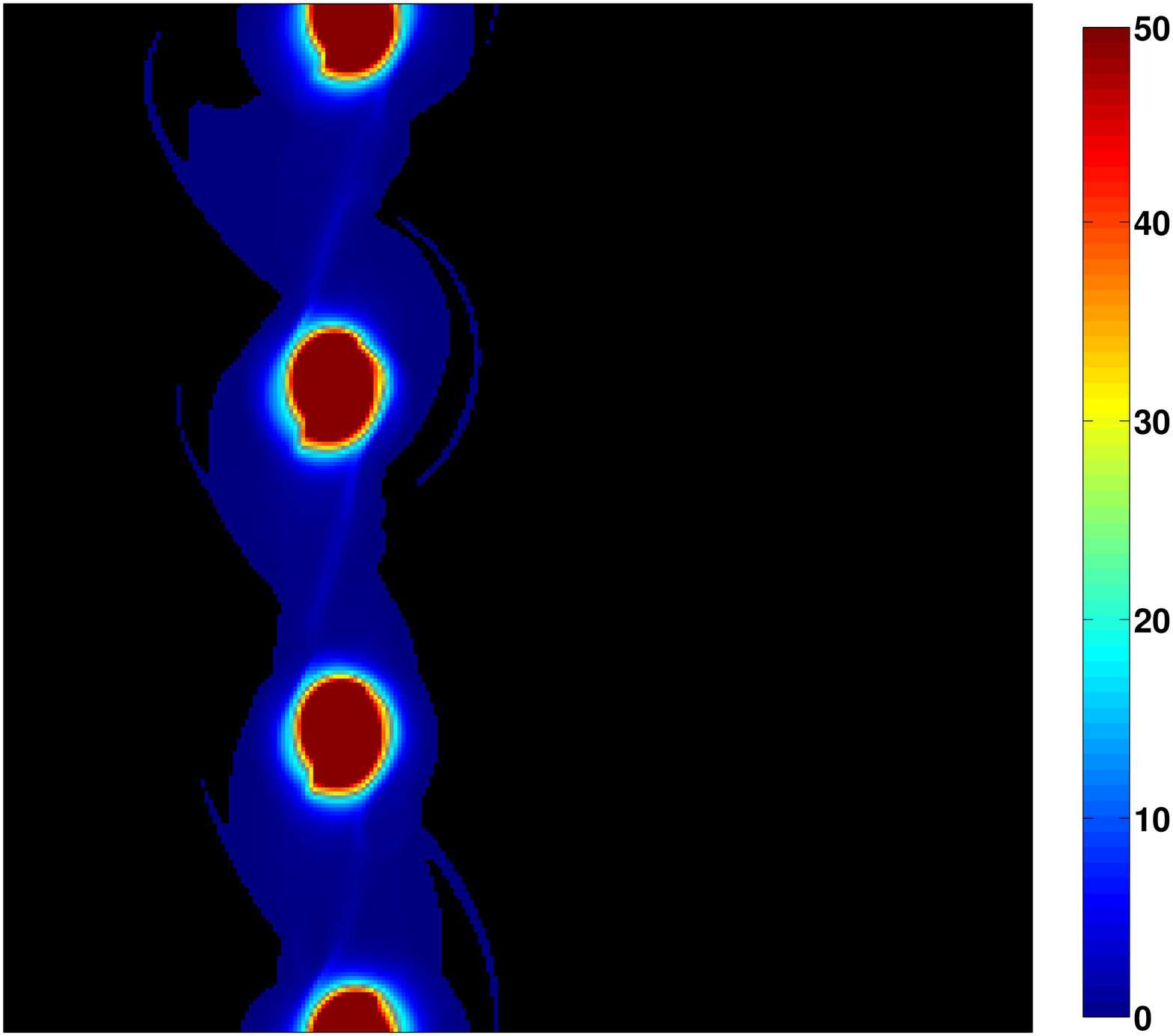} }
\hspace{5mm}
\subfigure[$t\Omega=60$]%[Surface density $\Sigma$: Red means $\Sigma\ge20\Sigma_0$
%and black means $\Sigma\approx0$($t=273\Omega^{-1}$).]
{\includegraphics[width=0.4\hsize]{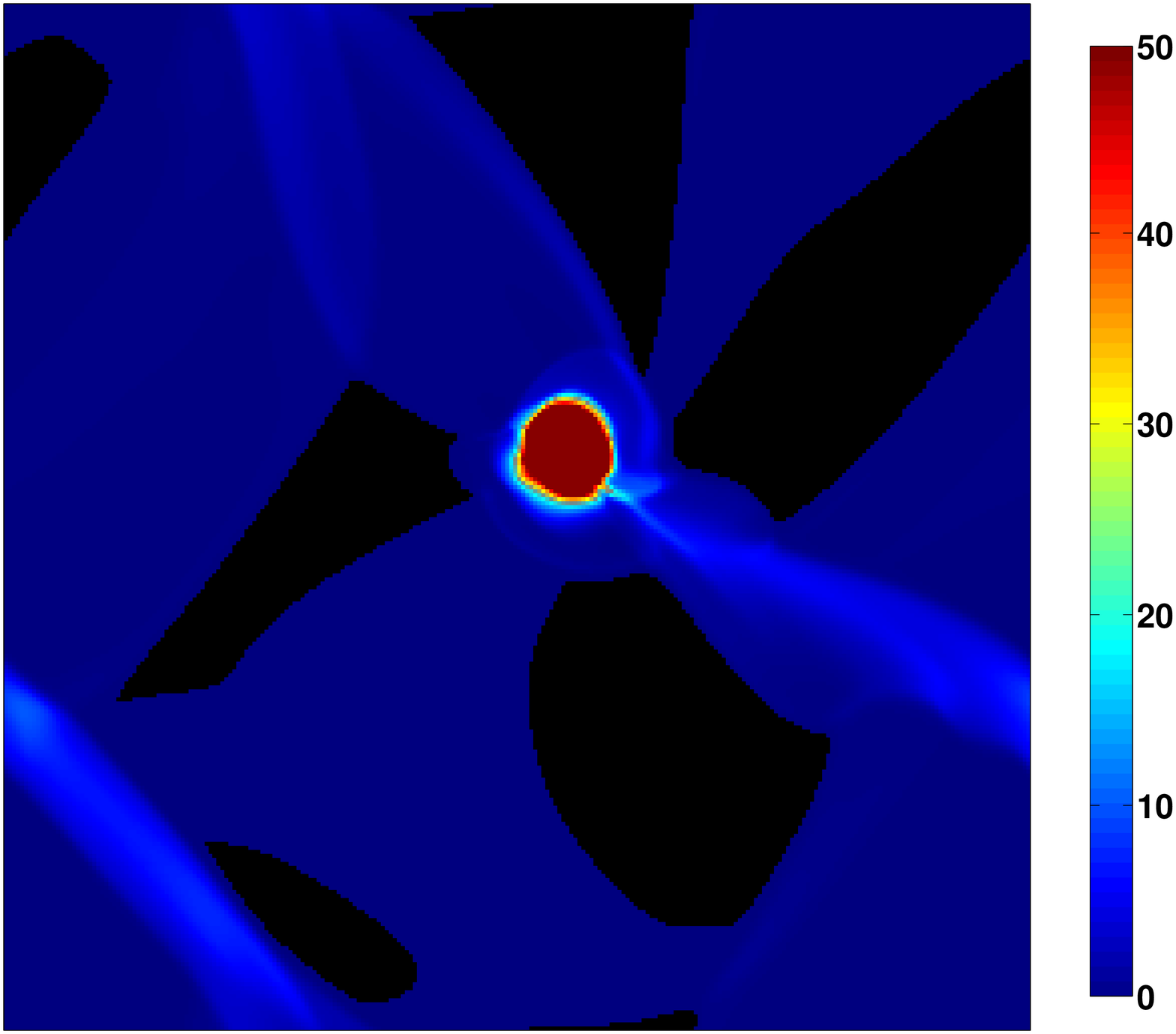}}
\caption{Simulation for $r_0=621r_s$, 
$\langle\Sigma\rangle_A=9.6\Sigma_0$ of $10^9\msun$ SMBH. {\it Line plots:}
Evolution history of accretion rate calculated from eq.~\eqref{mdot1},
$\alpha$, and mass-weighted $Q$ and $\beta$. Because the disk is not in thermal balance, 
the $\dot M$ shown here does not reflect that due to turbulent stress,
which would be smaller.
{\it Lower left:} Three fragments form at time $t=53\Omega^{-1}$. 
{\it Lower right:} A single fragment is left after mergers at time $t=60\Omega^{-1}$.}
\label{fragment_largemass}
\end{figure}

\subsection{Fragmentation at long cooling times}
\label{Results:CaseIII}
As noted in \S\ref{sec:intro}, the softening influence of radiation
pressure on the equation of state may allow fragmentation even for
$\Omega t_c\gg 1$.  One way to enter this regime is to increase the
mean surface density.  At fixed radius and fixed $Q$, the gas pressure
fraction $\beta\approx 0.5\hat\Sigma^{-3/2}$ [eq.~\ref{dimlessbetaQ}],
while the cooling time $t_c\propto\kappa\hat\Sigma^2$.  For example,
with $Q=1$ and $\hat{\Sigma}=6$ at $r_0=10^3r_s$ and $\mbh=10^8\msun$,
our equation of state yields $\beta=0.03$ and $\Omega t_{\rm c}=244$.
At $r_0\lesssim 200r_s$ along the fragmentation boundary shown in
Fig.~\ref{fragmentboundary}, $\beta$ is already very small and
$t_c\Omega$ is large.  The surface density exceeds what is required to
support the Eddington luminosity but might occur if mass
were dumped into the disk by a violent event such as a merger.

Another way to enter this regime is by increasing the mass of the
black hole at a fixed Eddington fraction, $\dot m$.  At
$\mbh=10^9\msun$ and $\dot m=1$, the disk is still strongly dominated
by radiation pressure at the smallest radii where self-gravity is important.
Setting $l_{\rm E}/\epsilon_{0.1}\equiv\dot m=1$, $M_8=10$, and
$\alpha=0.01$ in equations (A2)-(A4) of G03, we estimate that $Q=1$
occurs at $r_0=621r_s$, where $\beta=0.02$ and $\hat\Sigma=9.6$.

We have done one simulation with these values of $\mbh$, $r_0$, and
$\hat\Sigma$ (Figure~\ref{fragment_largemass}).  The simulation starts
at $Q=1.2$ and cooling time $t_c\Omega=49$. The large but declining
values of $\dot M$ in the first panel are computed from the thermal
equation~\eqref{mdot1} rather than the stress equation \eqref{mdot2},
which would predict $\dot M\approx 0$ in the initial phases when
$\alpha\approx 0$.  A sustained balance between radiative cooling and
turbulent heating is never achieved, even though the cooling is slow
and proceeds smoothly to $Q<1$.  After roughly one cooling time, at
$t\approx 50\Omega^{-1}$, the sheet collapses to an azimuthal
filament.  This quickly breaks into three massive fragments, which
merge into one at $60\Omega^{-1}$.  The energies of the final object
are $E_{\rm th}=4.6\times10^5E_0$, $E_{\rm kin}=7.5\times10^4E_0$,
$E_{\rm tid}=-15.1E_0$, and $E_{\rm grav}=-5.4\times10^5E_0$,
respectively, so that it is marginally bound.

As this example shows, the mass scale for fragmentation is very large
if it occurs at $\beta\ll 1$.  This is to be expected from the
Eddington quartic \eqref{Mbeta}.  The fragments inherit an initial
$\beta$ similar to that of the disk because their formation occurs
roughly adiabatically at high specific entropy,
$T^3/\rho\approx\mbox{constant}$.  The final object weighs
$893M_0\approx 9000\msun$ at $\langle\beta\rangle_m\approx 0.07$,
close to the prediction from eq.~\eqref{Mbeta} but larger than the
value $\beta_{\rm disk}\approx0.02$ expected for a uniform sheet at
$Q=1$ with this mean surface density.

We have not attempted a thorough a survey of parameter space for
$\mbh=10^9\msun$ as we did for $10^8\msun$.  One obstacle is that the
cooling time becomes very long, especially at small radii.  Another is
that the gravitational stress remains small up to the point of
fragmentation, in contrast to the situation for $10^8\mbh$ where
$\alpha_{\max}\approx 0.4$ (Fig.~\ref{fragmentboundary}).  At
$\alpha\lesssim 10^{-2}$, heating by MRI becomes important, which
cannot be explored with this 2D code.  MRI might have prevented
fragmentation for the disk parameters of
Fig.~\ref{fragment_largemass}, since $Q\ge 1$ is predicted for these
parameters if $\alpha\ge 10^{-2}$, and fragmentation did not begin
until $Q\lesssim0.5$. The point, however, is that self-gravity alone
was not able to supply a sufficiently large $\alpha$ despite the long
cooling time.  Thus our simulations demonstrate that fragmentation can
occur at dimensionless cooling times $\Omega t_{\rm c}\gg 1$ when
radiation pressure dominates, $\beta\ll 1$.

\section{Discussion and Conclusions}\label{sec:discussion}

Radiation pressure and appropriate opacities are part of the minimal
physics needed to study gravitational turbulence and fragmentation in
near-Eddington AGN accretion disks.  We have included these and used
them to test the predictions of Goodman (2003) for the maximum radius
at which AGN disks can support steady accretion at the Eddington rate
without fragmentation.  We are in qualitative agreement with that
paper, but quantitatively we find that the critical radius is about
twice as large as was predicted for $\mbh=10^8{\rm\,\msun}$.  We are
also generally in agreement with \citet{Gammie2001}'s criterion for
fragmentation, except that fragmentation may occur for $\Omega t_{\rm
  cool}\gg 1$ if radiation pressure dominates.

Beyond this, however, our local 2D approximation limits what we can
explore and what we can conclude.  Magnetohydrodynamic (MHD) and
magnetorotational (MRI) processes cannot be represented, at least not
directly.  This probably does not much alter the conditions for
marginal fragmentation, because the effective viscosity due to
self-gravity is much larger in this regime than what MRI can provide.
We cannot, however, exclude the possibility that MRI, or some other
effective viscosity that does not respect the conservation of specific
vorticity, might interact with the self-gravity in subtle ways, for
example by promoting secular instabilities at $Q>1$, or by enabling
transitions among states of different $\dot M$ at the same $\Sigma$
and $\Omega$; we saw evidence for the latter behavior when we added an
ad-hoc viscosity $\nu\propto P_{\rm gas}$ to our code, but we have not
been able to understand those results and therefore have not presented
them here.  Our approximations also cannot represent magnetized winds
or global spiral arms, which might in principle remove angular
momentum at rates enhanced by $\sim r/h$ compared to transport within
the disk, allowing a lower and hence less self-gravitating surface
density for the same accretion rate
\citep{Goodman2003,ThompsonQuataertMurray2005,HopkinsQuataert2010b}.

Nor can we test \cite{GoodmanTan2004}'s suggestion that fragments grow
rapidly up to the isolation mass, which is typically $\sim
10^5{\rm\,\msun}$.  There are two principal obstacles.  One is
numerical: very large self-gravitating masses are very strongly
radiation-pressure dominated, and therefore only marginally bound when
in virial equilibrium, so that small energy errors can cause spurious
expansion or collapse.  This is likely to be a difficulty for many
numerical algorithms besides ZEUS in low-$\beta$, self-gravitating
regimes.  The second is physical: our 2D results show that in shearing
sheets where multiple bound fragments are present, the gravitational
interactions between fragments quickly increases their epicyclic
motions to amplitudes exceeding their Hill radii, so that they would
be expected to scatter into the third dimension if that were allowed
\citep{Rafikov_Slepian2010}.

Notwithstanding these limitations of 2D, our results strongly suggest
that if the disks of bright QSOs extend at constant $\dot M$ to
$\gtrsim 0.01\mbox{-}0.1 {\rm\,pc}$, then bound objects will form with
individual masses of at least $300 \msun$, and possibly much more.
Collectively, these ``stars'' will dominate the local surface density
of the disk, though they may be accompanied by an optically thick
layer of distributed gas.  The stars will attain epicyclic dispersions
bounded by Safronov numbers $\Theta\equiv GM_*/(R_*\sigma_{r,\rm
  epi}^2)\lesssim 1$, so that even if they contract to their
main-sequence radii and are stable enough to remain there,
$(M_*/\mbh)(r/R_*)\ll 1$ unless $M_*\gtrsim 10^5 {\rm\,M_{\odot}}$
(see \citealt{GoodmanTan2004} for a review of the nominal
main-sequence properties of very massive stars).  Physical collisions
will be important unless or until the objects collapse to black holes.
One is thus lead to imagine a model for the disk similar to that
advanced long ago by \citet{Spitzer_Saslaw1966}, and more recently by
\citet{Miralda-Escude_Kollmeier2006}.  We imagine formation
of (very massive) stars within a disk, however, rather than formation of a disk
from a pre-existing dense nuclear star cluster.

Are there any observational signatures of such a fragmented disk that
might distinguish it from the conventionally imagined smooth one?  One
such signature may be the super-solar metallicity inferred from the
broad lines, which appears not to correlate with the general star
formation rate in the host traced by far-infrared emission
\citep{Simon_Hamann2010}, and therefore may implicate formation within
the nuclear disk itself.  Another signature might be deviations from
the spectral energy distribution expected from a steadily accreting,
optically thick disk.  \cite{GoodmanTan2004} pointed out that the
viscous accretion time at $r\sim 10^3\rS$ is typically somewhat less
than the minimum main-sequence lifetime, so that massive stars formed
there might---if they are sufficiently stable---migrate to the inner
edge of the disk before dying.  In that case, if these stars dominate
the surface density and are not fully enshrouded by diffuse gas, the
local color temperature of the disk might be intermediate between that of the
stars themselves and the effective temperature implied by the
accretion rate.  Gravitational microlensing observations are beginning
to test the variation of apparent disk size with wavelength on
relevant lengthscales; the evidence is consistent with color
temperature $\propto r^{-3/4}$ as expected for steady, optically thick disks, but suggests that
the disks are larger at a given wavelength than expected
\citep{Pooley_etal2007,Morgan_etal2010}.

%Finally, here we fix the gas metallicity in the disk to be
%the solar value. If the abundance is larger than this value as
%suggested by some observations (REFERENCEs), at a fix radius,
%Krammer's opacity would be larger and the cooling time would be
%longer. To get the same cooling effect we have to go to larger
%radii. We have done simulations with Krammer's opacity increased by
%a factor of 10. In those simulations for one Eddington accretion
%rate, the boundary between fragmentation and no fragmentation moves
%to about $5\times10^3r_s$. So the change of metallicity would not
%change our results significantly.

% Recently, based on eleven gravitationally lensed quasars observed 
% with microlensing variability, \cite{Morgan_etal2010} gives a 
% relation between the accretion disk size and black hole mass, which is 
% consistent with thin disk theory. However, they find that accretion 
% disk must have flatter temperature profile compared with standard 
% thin disk model.  According to our results, accretion disk is already 
% not in a uniform accretion state to $\sim10^3r_s$. There is large density 
% fluctuation and fragmentation to $4\times10^3r_s$. 
% When we observe the fluctuate and clumpy disk, 
% the average temperature we get is higher than what we expect 
% from uniform standard think disk. Then this will make the temperature 
% profile flatter compared with standard thin disk theory.   

\section*{ACKNOWLEDGEMENT}
We thank Charles F. Gammie to give us his code and helpful comments to
run the code. Yan-Fei Jiang thanks Jim Stone for helpful discussion on
numerical issue on the code. Yan-Fei Jiang also thanks Jerry Ostriker,
Renyue Cen for helpful discussions.  This work was supported in part
by the NSF Center for Magnetic Self-Organization under NSF grant
PHY-0821899.

\begin{appendix}
\section{Effective 2D Equation of State for vertically constant $\beta$}
Vertical hydrostatic equilibrium in the combined gravitational fields
of the central mass and of the disk itself is described by
\begin{equation}
\frac{1}{\rho}\frac{dp}{dz}=-\Omega^2z-4\pi
G\int_0^{z}\rho(z^{\prime})dz^{\prime}\,,\label{verticalbalance}
\end{equation}
Putting $p=K(\beta)\rho^{4/3}$ [eq.~\eqref{pressureK}]
and adopting the dimensionless Emden variables
\begin{equation}
\rho(z)=\rho(0)\theta^3,\qquad \xi=z/h,\qquad h^2=\frac{K}{\pi
G\rho(0)^{2/3}}\,,\label{Emden}
\end{equation}
leads to
\begin{equation}
\frac{d^2\theta}{d\xi^2}+\theta^3=-\frac{\Omega^2}{4\pi
G\rho(0)}\equiv-\frac{Q}{2}\,. \label{dimlessbalance}
\end{equation}
Using the initial conditions $\theta(0)=1$ and $\theta'(0)=0$, eq.~\eqref{dimlessbalance}
can be reduced to a quadrature:
\begin{equation}
\xi=2\sqrt{2}\int_0^{\sqrt{1-\theta}}\frac{dw}{\sqrt{2Q+1+(1-w^2)+(1-W^2)^2+(1-w^2)^3}}.
\end{equation}
The height-integrated density and pressure become
\begin{equation}
\Sigma=4\sqrt{2}\rho(0)hI_3(Q),\qquad P=4\sqrt{2}K(\beta)\rho(0)^{4/3}h I_4(Q),
\label{SigmaP}
\end{equation}
in which
\begin{equation}
I_k(Q)\equiv
\int_0^1\frac{(1-w^2)^kdw}{\sqrt{2Q+1+(1-w^2)+(1-w^2)^2+(1-w^2)^3}},\qquad
k=3,4. \label{fullIkQ}
\end{equation}

For all $w\in[0,1]$ and $Q\ge0$,  the denominators of the
elliptic integrals (\ref{fullIkQ}) vary by at most a factor of 2.
Hence we approximate these integrals with
single-point Gaussian quadrature scheme,
\begin{equation}
\int_0^1(1-w^2)^k f(w^2)dk\approx u_k(w_k^2),
\end{equation}
in which the point $w_k\in[0,1]$ and the weight $u_k>0$ are chosen so
as to make this scheme exact for functions $f(w^2)=A+Bw^2$ with
arbitrary constants $A$ and $B$. For $k=3$ and $k=4$, the two
integrals are close enough that the same value of $w_k$ can be used
for both; this leads to the approximations (\ref{I3I4}), which are
accurate to $\lesssim 1\%$ for all $Q\ge0$.  In practice, we prepare a
table with a certain range of $\Sigma$ and $U$, within which we
calculate the integrals \eqref{fullIkQ} directly.  For
conditions outside the table, the code uses the
approximations \eqref{I3I4}.

\begin{figure}
\centering
%\subfigure[]{\includegraphics[width=0.46\hsize]{Q_Sigma} }
\subfigure[]{\includegraphics[width=0.46\hsize]{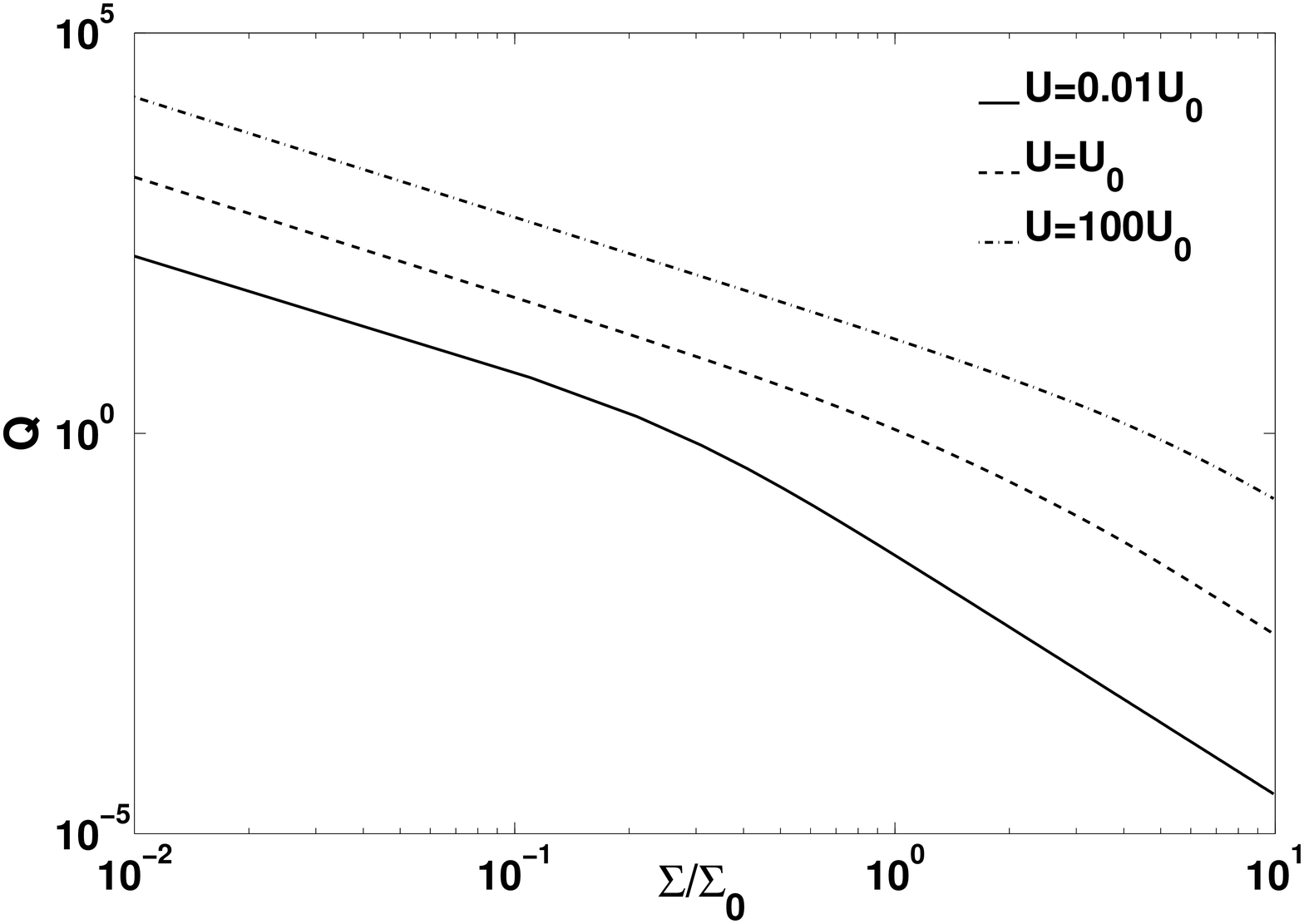} }
\hspace{5mm}
\subfigure[]{\includegraphics[width=0.46\hsize]{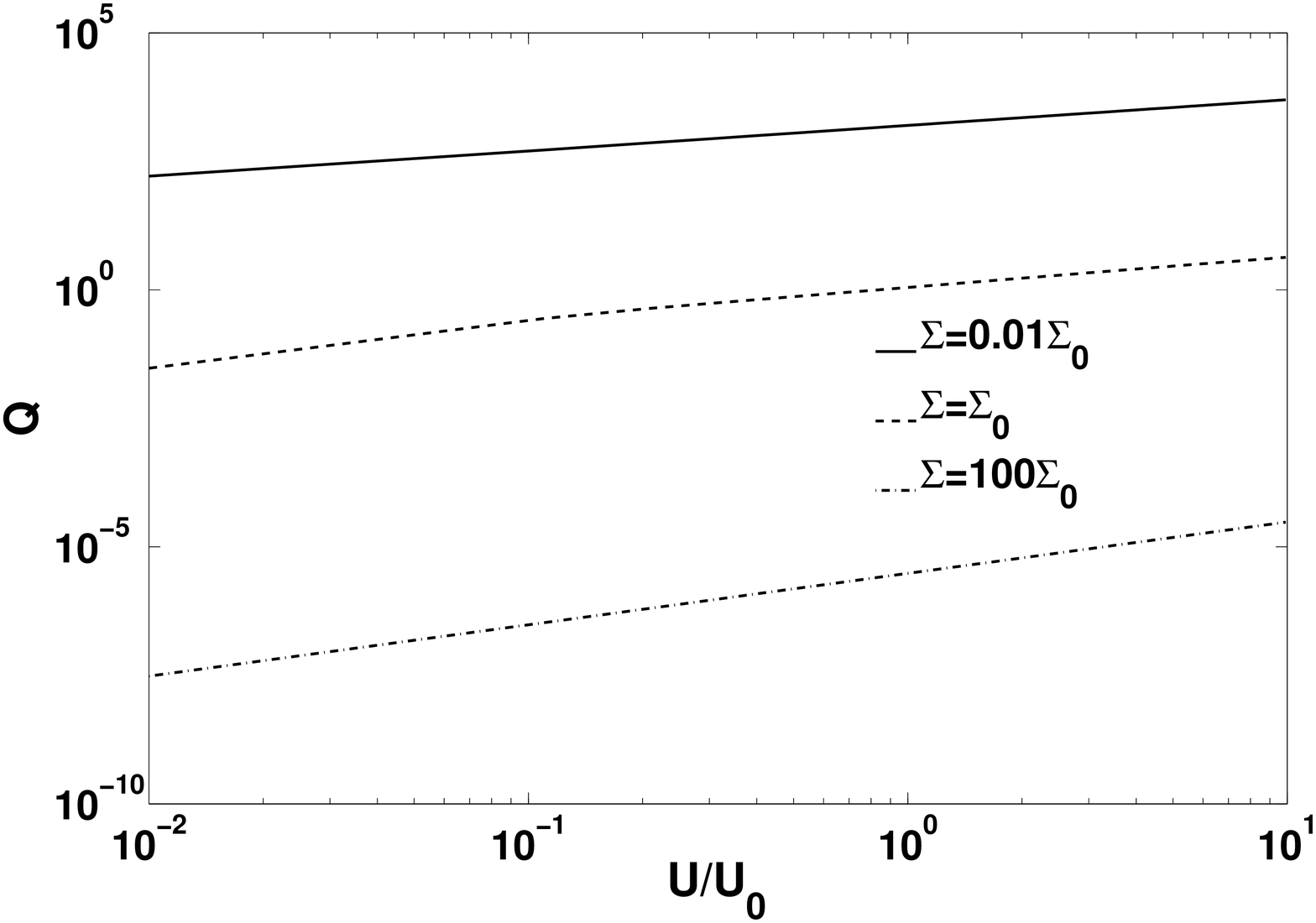} }\\
%\subfigure[]{\includegraphics[width=0.46\hsize]{beta_Sigma}}
\subfigure[]{\includegraphics[width=0.46\hsize]{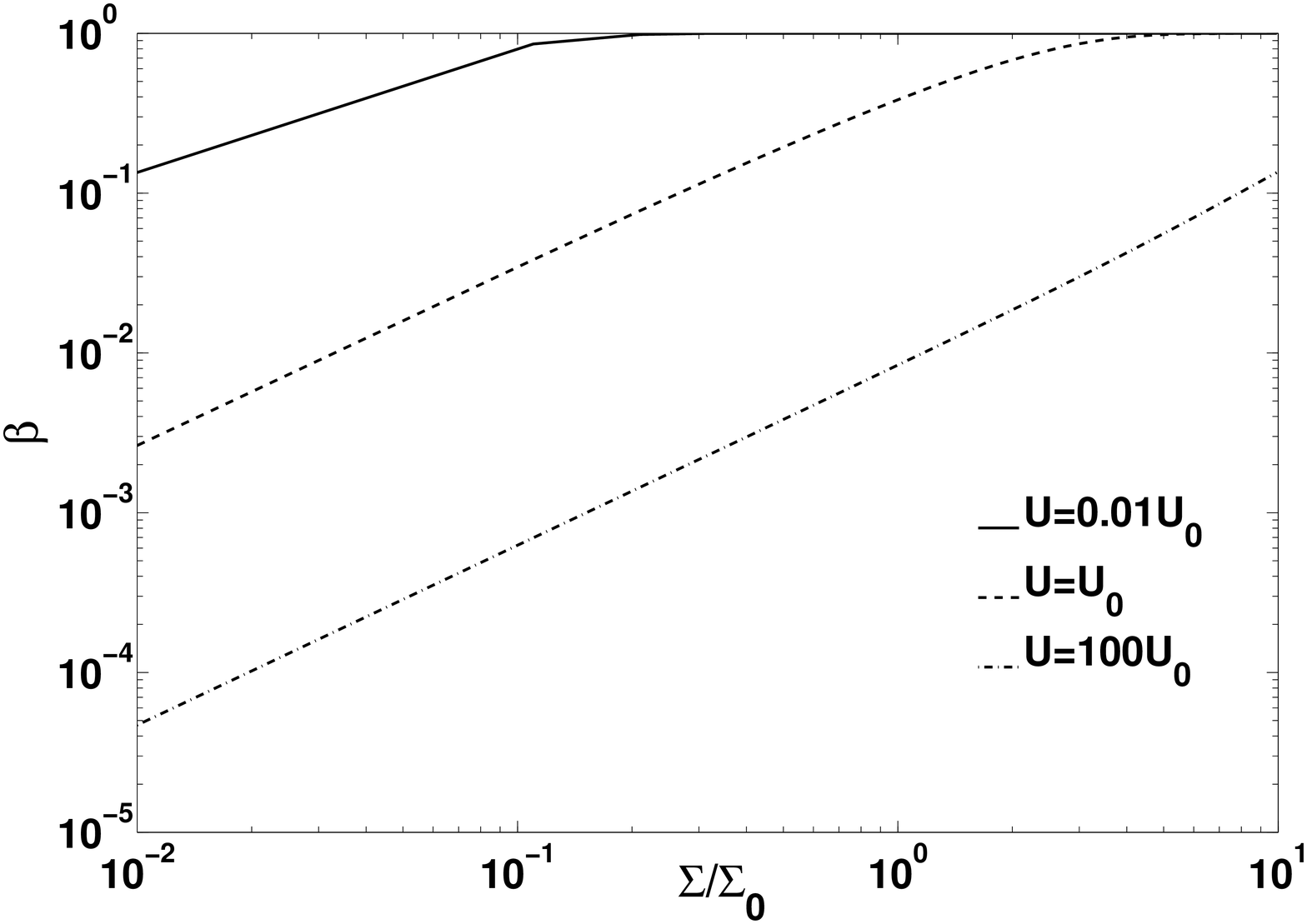}}
\hspace{5mm}
\subfigure[]{\includegraphics[width=0.46\hsize]{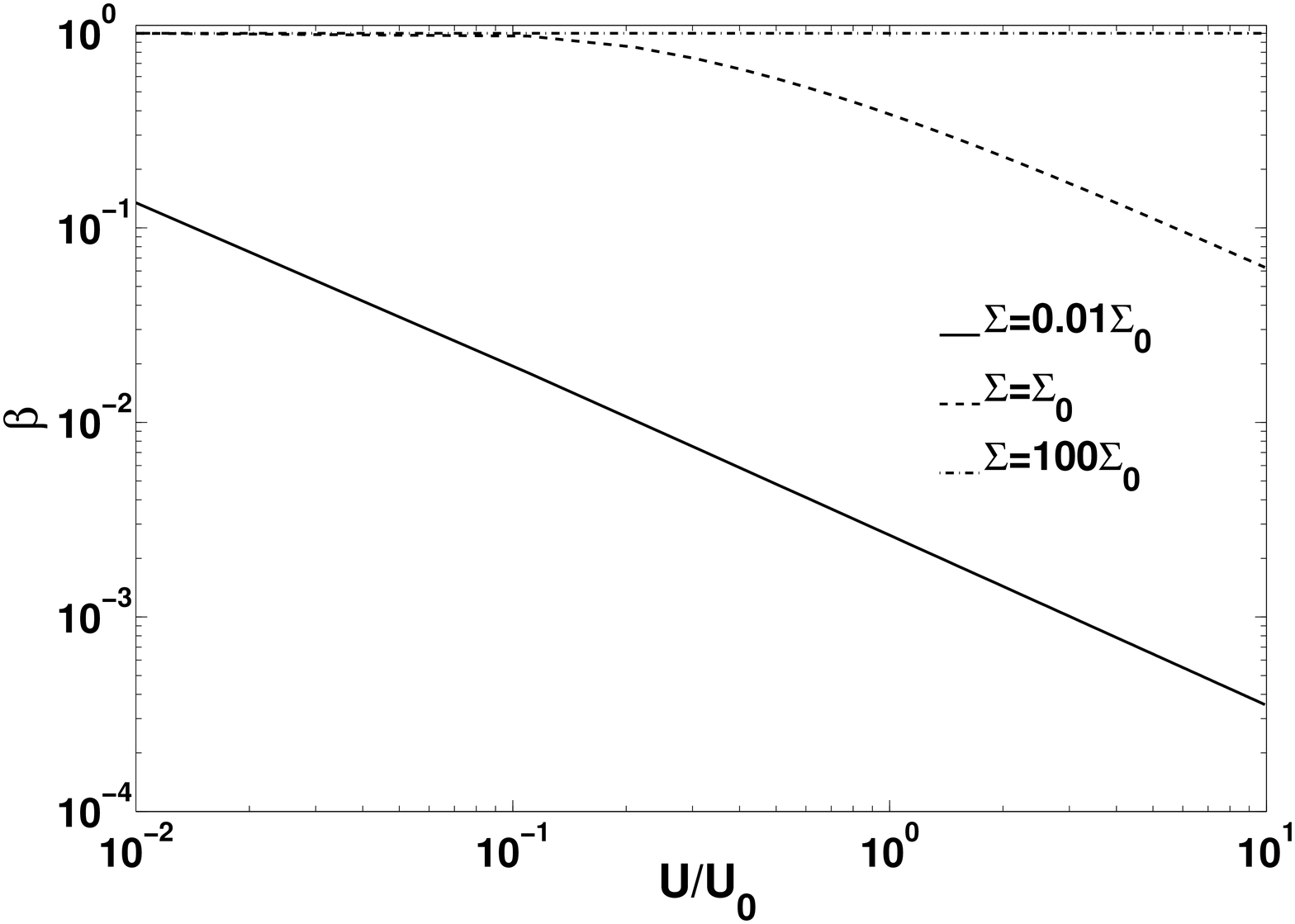}}\\
\caption{Solutions to equation (\ref{dimlessbetaQ}) for $\beta$ and
$Q$ in the parameter space $\hat{\Sigma}$ and $\hat{U}$. In panel
(a) and (c), we show the behavior of $Q$ and $\beta$ as a function
of $\hat{\Sigma}$ for a fixed value of $\hat{U}$. Different lines
are for different values of $\hat{U}$ as shown in the plots. In
panel (b) and (d), we show the behavior of $Q$ and $\beta$ as a
function of $\hat{U}$ for a fixed value of $\hat{\Sigma}$. The
values of $\hat{\Sigma}$ for different lines are also shown in the
plots. The four plots are the equation of state that we use in our
code. } \label{Q_Beta_S_U}
\end{figure}

Eliminating $h$ between eqs.~\eqref{Emden} and the first of
eqs.~\eqref{SigmaP} and then expressing $\rho(0)$ in terms of $Q$ via eq.~\eqref{Q}
leads to eq.~\eqref{betaandQ}.
Using this to eliminate $K$ and $h$ from the second of
eqs.~\eqref{SigmaP} then yields equation \eqref{EOS} for $P$ in terms
of $Q$ and $\Sigma$.  But $P$ is related to the internal energy by
eq.~\eqref{UvsP}, so \eqref{EOS} can be recast as \eqref{EOS2}.
Finally, after replacing $K$ with its explicit form \eqref{pressureK},
equations \eqref{betaandQ} and \eqref{EOS2}
can be rewritten in terms of the dimensionless variables introduced in
\S\ref{sec:units} as
\begin{equation}
\frac{2^{15}(1-\beta)}{\beta^4}=\frac{\pi^3Q^4}{\left[I_3(Q)\right]^6}\hat{\Sigma}^6,\qquad
\beta=2-\frac{128\left[I_3(Q)\right]^3}{3QI_4(Q)}\frac{\hat{U}}{\hat{\Sigma}^3}.
\label{dimlessbetaQ}
\end{equation}
Equations \eqref{dimlessbetaQ} implicitly determine $\beta$ and $Q$ given
$\hat U$ and $\hat\Sigma$, as exemplified by Fig.~\ref{Q_Beta_S_U}, after which $P$ follows from
eqs.~\eqref{UvsP} or \eqref{EOS}.
%If $\beta\ll 1$, then $\beta\propto\hat{\Sigma}^{-3/2}$ at fixed $Q$.

\end{appendix}

\bibliographystyle{apj}
\bibliography{QuasarDisk}

\begin{thebibliography}{43}
\expandafter\ifx\csname natexlab\endcsname\relax\def\natexlab#1{#1}\fi

\bibitem[{{Bartko} {et~al.}(2010)}]{Bartko_etal2010}
{Bartko}, H., {et~al.} 2010, \apj, 708, 834

\bibitem[{{Bond} {et~al.}(1984){Bond}, {Arnett}, \&
  {Carr}}]{BondArnettCarr1984}
{Bond}, J.~R., {Arnett}, W.~D., \& {Carr}, B.~J. 1984, \apj, 280, 825

\bibitem[{{Collin} \& {Zahn}(1999)}]{CollinZahn1993}
{Collin}, S., \& {Zahn}, J. 1999, \aap, 344, 433

\bibitem[{{Davies} {et~al.}(2007){Davies}, {S{\'a}nchez}, {Genzel}, {Tacconi},
  {Hicks}, {Friedrich}, \& {Sternberg}}]{Daviesetal2007}
{Davies}, R.~I., {S{\'a}nchez}, F.~M., {Genzel}, R., {Tacconi}, L.~J., {Hicks},
  E.~K.~S., {Friedrich}, S., \& {Sternberg}, A. 2007, \apj, 671, 1388

\bibitem[{{Dhanda} {et~al.}(2007){Dhanda}, {Baldwin}, {Bentz}, \&
  {Osmer}}]{Dhanda_etal2007}
{Dhanda}, N., {Baldwin}, J.~A., {Bentz}, M.~C., \& {Osmer}, P.~S. 2007, \apj,
  658, 804

\bibitem[{{Gammie}(2001)}]{Gammie2001}
{Gammie}, C.~F. 2001, \apj, 553, 174

\bibitem[{{Ghez} {et~al.}(2003){Ghez}, {Duch{\^e}ne}, {Matthews}, {Hornstein},
  {Tanner}, {Larkin}, {Morris}, {Becklin}, {Salim}, {Kremenek}, {Thompson},
  {Soifer}, {Neugebauer}, \& {McLean}}]{Ghez_etal2003}
{Ghez}, A.~M., {et~al.} 2003, \apjl, 586, L127

\bibitem[{{Goodman}(2003)}]{Goodman2003}
{Goodman}, J. 2003, \mnras, 339, 937, {\bf (G03)}

\bibitem[{{Goodman} \& {Tan}(2004)}]{GoodmanTan2004}
{Goodman}, J., \& {Tan}, J.~C. 2004, \apj, 608, 108

\bibitem[{{Hamann} \& {Ferland}(1993)}]{HamannFerland1993}
{Hamann}, F., \& {Ferland}, G. 1993, \apj, 418, 11

\bibitem[{{Hawley} {et~al.}(1995){Hawley}, {Gammie}, \&
  {Balbus}}]{HawleyGammieBalbus1995}
{Hawley}, J.~F., {Gammie}, C.~F., \& {Balbus}, S.~A. 1995, \apj, 440, 742

\bibitem[{{Hirose} {et~al.}(2009){Hirose}, {Blaes}, \&
  {Krolik}}]{HiroseBlaesKrolik2009}
{Hirose}, S., {Blaes}, O., \& {Krolik}, J.~H. 2009, \apj, 704, 781

\bibitem[{{Hopkins} \& {Quataert}(2010{\natexlab{a}})}]{HopkinsQuataert2010b}
{Hopkins}, P.~F., \& {Quataert}, E. 2010{\natexlab{a}}, ArXiv e-prints

\bibitem[{{Hopkins} \& {Quataert}(2010{\natexlab{b}})}]{HopkinsQuataert2010a}
---. 2010{\natexlab{b}}, \mnras, 1085

\bibitem[{{Hubeny}(1990)}]{Hubeny1990}
{Hubeny}, I. 1990, \apj, 351, 632

\bibitem[{{Johnson} \& {Gammie}(2003)}]{JohnsonGammie2003}
{Johnson}, B.~M., \& {Gammie}, C.~F. 2003, \apj, 597, 131

\bibitem[{{Kuo} {et~al.}(2010){Kuo}, {Braatz}, {Condon}, {Impellizzeri}, {Lo},
  {Zaw}, {Schenker}, {Henkel}, {Reid}, \& {Greene}}]{Kuo_etal2010}
{Kuo}, C.~Y., {et~al.} 2010, ArXiv e-prints

\bibitem[{{Lauer} {et~al.}(2005){Lauer}, {Faber}, {Gebhardt}, {Richstone},
  {Tremaine}, {Ajhar}, {Aller}, {Bender}, {Dressler}, {Filippenko}, {Green},
  {Grillmair}, {Ho}, {Kormendy}, {Magorrian}, {Pinkney}, \&
  {Siopis}}]{Laueretal2005}
{Lauer}, T.~R., {et~al.} 2005, \aj, 129, 2138

\bibitem[{{Lee} \& {Goodman}(1999)}]{LeeGoodman1999}
{Lee}, E., \& {Goodman}, J. 1999, \mnras, 308, 984

\bibitem[{{Lightman} \& {Eardley}(1974)}]{LightmanEardley1974}
{Lightman}, A.~P., \& {Eardley}, D.~M. 1974, \apjl, 187, L1+

\bibitem[{{Lodato} \& {Rice}(2004)}]{LodatoRice2004}
{Lodato}, G., \& {Rice}, W.~K.~M. 2004, \mnras, 351, 630

\bibitem[{{Martins} {et~al.}(2008){Martins}, {Gillessen}, {Eisenhauer},
  {Genzel}, {Ott}, \& {Trippe}}]{Martins_etal2008}
{Martins}, F., {Gillessen}, S., {Eisenhauer}, F., {Genzel}, R., {Ott}, T., \&
  {Trippe}, S. 2008, \apjl, 672, L119

\bibitem[{{Masset}(2000)}]{Masset2000}
{Masset}, F. 2000, \aaps, 141, 165

\bibitem[{{Miller} \& {Antonucci}(1983)}]{MillerAntonucci1983}
{Miller}, J.~S., \& {Antonucci}, R.~R.~J. 1983, \apjl, 271, L7

\bibitem[{{Miralda-Escud{\'e}} \&
  {Kollmeier}(2006)}]{Miralda-Escude_Kollmeier2006}
{Miralda-Escud{\'e}}, J., \& {Kollmeier}, J.~A. 2006, \nar, 50, 786

\bibitem[{{Miyoshi} {et~al.}(1995){Miyoshi}, {Moran}, {Herrnstein},
  {Greenhill}, {Nakai}, {Diamond}, \& {Inoue}}]{Miyoshi_etal1995}
{Miyoshi}, M., {Moran}, J., {Herrnstein}, J., {Greenhill}, L., {Nakai}, N.,
  {Diamond}, P., \& {Inoue}, M. 1995, \nat, 373, 127

\bibitem[{{Monaghan} \& {Price}(2001)}]{MonaghanPrice2001}
{Monaghan}, J.~J., \& {Price}, D.~J. 2001, \mnras, 328, 381

\bibitem[{{Morgan} {et~al.}(2010){Morgan}, {Kochanek}, {Morgan}, \&
  {Falco}}]{Morgan_etal2010}
{Morgan}, C.~W., {Kochanek}, C.~S., {Morgan}, N.~D., \& {Falco}, E.~E. 2010,
  \apj, 712, 1129

\bibitem[{{Nayakshin} \& {Sunyaev}(2005)}]{NayakshinSunyaev2005}
{Nayakshin}, S., \& {Sunyaev}, R. 2005, \mnras, 364, L23

\bibitem[{{Pooley} {et~al.}(2007){Pooley}, {Blackburne}, {Rappaport}, \&
  {Schechter}}]{Pooley_etal2007}
{Pooley}, D., {Blackburne}, J.~A., {Rappaport}, S., \& {Schechter}, P.~L. 2007,
  \apj, 661, 19

\bibitem[{{Pringle}(1981)}]{Pringle1981}
{Pringle}, J.~E. 1981, \araa, 19, 137

\bibitem[{{Rafikov} \& {Slepian}(2010)}]{Rafikov_Slepian2010}
{Rafikov}, R.~R., \& {Slepian}, Z.~S. 2010, \aj, 139, 565

\bibitem[{{Rice} {et~al.}(2003){Rice}, {Armitage}, {Bate}, \&
  {Bonnell}}]{RiceArmitageBonnell2003}
{Rice}, W.~K.~M., {Armitage}, P.~J., {Bate}, M.~R., \& {Bonnell}, I.~A. 2003,
  \mnras, 339, 1025

\bibitem[{{Rice} {et~al.}(2005){Rice}, {Lodato}, \&
  {Armitage}}]{RiceLodatoArmitage2005}
{Rice}, W.~K.~M., {Lodato}, G., \& {Armitage}, P.~J. 2005, \mnras, 364, L56

\bibitem[{{Shlosman} {et~al.}(1990){Shlosman}, {Begelman}, \&
  {Frank}}]{Shlosmanetal1990}
{Shlosman}, I., {Begelman}, M.~C., \& {Frank}, J. 1990, \nat, 345, 679

\bibitem[{{Simon} \& {Hamann}(2010)}]{Simon_Hamann2010}
{Simon}, L.~E., \& {Hamann}, F. 2010, \mnras, 407, 1826

\bibitem[{{So{\l}tan}(1982)}]{Soltan1982}
{So{\l}tan}, A. 1982, \mnras, 200, 115

\bibitem[{{Spitzer} \& {Saslaw}(1966)}]{Spitzer_Saslaw1966}
{Spitzer}, Jr., L., \& {Saslaw}, W.~C. 1966, \apj, 143, 400

\bibitem[{{Springel} {et~al.}(2005){Springel}, {Di Matteo}, \&
  {Hernquist}}]{SpringelDiMatteoHernquist2005}
{Springel}, V., {Di Matteo}, T., \& {Hernquist}, L. 2005, \mnras, 361, 776

\bibitem[{{Stone} \& {Norman}(1992)}]{StoneNorman1992}
{Stone}, J.~M., \& {Norman}, M.~L. 1992, \apjs, 80, 753

\bibitem[{{Thompson} {et~al.}(2005){Thompson}, {Quataert}, \&
  {Murray}}]{ThompsonQuataertMurray2005}
{Thompson}, T.~A., {Quataert}, E., \& {Murray}, N. 2005, \apj, 630, 167

\bibitem[{{Toomre}(1964)}]{Toomre1964}
{Toomre}, A. 1964, \apj, 139, 1217

\bibitem[{{Yu} \& {Tremaine}(2002)}]{YuTremaine2002}
{Yu}, Q., \& {Tremaine}, S. 2002, \mnras, 335, 965

\end{thebibliography}

\end{document}